\documentclass[12pt]{article}
\usepackage{latexsym}
\usepackage{comment}
\usepackage{amsthm}
\usepackage{verbatim,color}
\usepackage{graphicx}
\usepackage{amssymb}
\usepackage{amssymb, color}
\usepackage{amsmath}
\usepackage{latexsym}
\usepackage{wrapfig}
\usepackage{subfigure}
\usepackage{wasysym}

\usepackage[utf8]{inputenc}
\usepackage{multicol}
\usepackage{url}
\usepackage{siunitx}
\usepackage{hyperref}
\usepackage{doi}
\usepackage{makecell}
\usepackage{tikz}
\usetikzlibrary{shapes}
\usepackage{pifont}

\makeatletter
\def\blfootnote{\xdef\@thefnmark{}\@footnotetext}
\makeatother
\newcommand{\R}{\mathbb{R}}
\newcommand{\Z}{\mathbb{Z}}
\newcommand{\bfa}{{\bf a}}
\newcommand{\bfb}{{\bf b}}
\newcommand{\bfc}{{\bf c}}

\newcommand{\bfe}{{\bf e}}

\newcommand{\bfn}{{\bf n}}

\newcommand{\bfp}{{\bf p}}

\newcommand{\bfr}{{\bf r}}

\newcommand{\bft}{{\bf t}}

\newcommand{\bfx}{{\bf x}}
\newcommand{\bfy}{{\bf y}}

\newcommand{\bfz}{{\bf z}}

\newcommand{\bfC}{{\bf C}}

\newcommand{\bfI}{{\bf I}}

\newcommand{\bfQ}{{\bf Q}}
\newcommand{\bfR}{{\bf R}}

\newcommand{\bfT}{{\bf T}}
\newcommand{\bfU}{{\bf U}}

\newcommand{\bfX}{{\bf X}}

\newcommand{\beq}{\begin{equation}}
\newcommand{\eeq}{\end{equation}}
\newcommand{\beqs}{\begin{eqnarray}}
\newcommand{\eeqs}{\end{eqnarray}}

\newcommand{\calC}{{\cal C}}

\newcommand{\calG}{{\cal G}}

\newcommand{\calI}{{\cal I}}
\newcommand{\calL}{{\cal L}}
\newcommand{\calM}{{\cal M}}

\newcommand{\calR}{{\cal R}}
\newcommand{\calS}{{\cal S}}
\newcommand{\calT}{{\cal T}}
\newcommand{\calU}{{\cal U}}

\newtheorem{theorem}{Theorem}[section]

\newtheorem{proposition}{Proposition}[section]

\newcommand{\bfig}{\begin{figure}[!h]}	
\newcommand{\efig}{\end{figure}}		

\begin{document}

\begin{center}
\Huge

{\bf Origami and materials science} \\

\vspace{5mm}
\normalsize

\vspace{2mm}
Huan Liu$^{1}$, Paul Plucinsky$^{2}$, Fan Feng$^{3}$ and Richard D. James$^{1}$

\vspace{2mm}

\vspace{2mm}
$^{1}$ Department of Aerospace Engineering and Mechanics,
University of Minnesota, Minneapolis, MN 55455, USA  \ \ 
liu01003@umn.edu, james@umn.edu \\
$^{2}$Department of Aerospace and Mechanical Engineering,
University of Southern California,  Los Angeles, CA 90089, USA \ \ 
 plucinsk@usc.edu \\
$^{3}$Cavendish Laboratory, University of Cambridge, Cambridge CB3 0HE, UK \ \ 
ff342@cam.ac.uk \\

\end{center}

\vspace{0.2in}
\normalsize

\noindent {\normalsize {\bf Abstract.}  Origami, the ancient art of folding thin sheets, has
attracted increasing attention for its practical value in diverse fields:
architectural design, therapeutics, deployable space structures, medical stent design, antenna design and
robotics.  In this survey article we highlight its suggestive value for the design of materials.
At continuum level the rules for constructing origami have direct analogs in the analysis
of the microstructure of materials. At atomistic level the structure of crystals, nanostructures,
viruses and quasicrystals all link to simplified methods of constructing origami.  
Underlying these linkages are basic physical scaling laws, the role of isometries,
and the simplifying role of group theory.  Non-discrete isometry groups suggest an unexpected framework 
for the  possible design of novel materials.} 

\tableofcontents

\vspace{0.2in}

\vspace{2mm}

\section{Introduction: the periodic table and objective structures}
\label{sect1}

In this article we collect together some surprising links between methods for the 
construction of origami structures and strategies for the design of materials.  The presentation
is nontechnical and draws from recent papers on both subjects, while forging new links that
were not developed or explained in detail.

From the perspective of the design of materials,  origami connects closely with the viewpoint of 
objective structures \cite{james2006objective} (defined below).  In the simplest case  one can think of the periodic table.  
As a way of quantifying the structure of materials, the conventional method \cite{hahn} is via crystal structure, i.e., 
the face-centered cubic (FCC) and  body-centered cubic (BCC) Bravais lattices that make up over half
the periodic table, together with the non-Bravais lattices such as HCP and the diamond structure.  Here, in considering
of  the periodic table we consider only the stable elements, i.e., the first 6 rows\footnote{excluding Astatine, whose structure
is not known} and we use the structure at room temperature if it is solid; otherwise, we use the accepted crystal
structure at $\approx0$ K. 

From the viewpoint of objective structures  the environment seen by an atom, rather than how the
atoms are arranged in space, is the basic concept.  In the
simplest case of the elements,  considers an atomic structure $\calS = \{ \bfx_i \in \R^3: i = 1, 2, \dots, N\}$
where $N \le \infty$.  We say that it is an {\it objective atomic structure} if $\calS$ is discrete and,
for each $i = 1, 2, \dots, N$, there is an orthogonal tensor $\bfQ_i$ such that
\beq
\{ \bfQ_i(\bfx_j - \bfx_1) + \bfx_i: j = 1, \dots, N \} = \calS,  \label{oas}
\eeq
i.e., each atom sees the same environment up to orthogonal transformation.  As described in \cite{james2018symmetry},
the structures of elements in the first six rows of the periodic table, including Bravais and non-Bravais lattices and 
structures that are not lattices at all, comply with (\ref{oas}), with few counterexamples.  Also included are the
celebrated forms of carbon: carbon nanotubes (any chirality), graphene, and buckminsterfullerine (C$_{60}$).
A glaring counter-example is manganese.  In fact, bulk manganese, whose structure is the union of four interpenetrating
Bravais lattices is better considered as an alloy than an element, due to degenerate spin configurations \cite{hobbs_03}.

This concept (\ref{oas}) could apply to the vertices of an origami structure, and we use this interpretation in some 
of the examples below.
An alternative concept, also used below, will be one that applies to the tiles, that is, to collections of points.  
The atomistic analog of a tile is a molecule. In an (ideal) origami structure each point on a tile is labelled by $\bfx \in \calT $ 
in the flat configuration (before folding), where the tile $\calT \subset \R^2$ is a connected region 
bounded by creases.  

The analog of (\ref{oas}) for a collection of molecules is a set of points 
$\calS = \{ \bfx_{i,j}:  i = 1, \dots, N,\ \ j = 1, \dots, M \}$ where $N \le \infty$ and $M < \infty$, i.e., $N$ molecules, 
each with $M$ atoms.  Here, consistent with (\ref{oas}),  $\bfx_{i,j}$ represents the position of atom $j$ of molecule $i$.  
A useful generalization of an objective atomic structure to molecules is that {\it corresponding atoms in different molecules see the same 
environment}.  We can renumber the atoms within a molecule so that ``corresponding'' means having the same 
index $j$.  Then corresponding atoms see the same environment if, for each $i = 1, \dots, N,\ \ j = 1, \dots, M $,
there is an orthogonal tensor $\bfQ_{i,j}$ depending in general on both $i$ and $j$ such that
\beq
\{ \bfQ_{i,j}(\bfx_{p,q} - \bfx_{1,j}) + \bfx_{i,j}: p = 1, \dots, N, \ \  q = 1, \dots, M \} = \calS,  \label{oms}
\eeq
If so, we call $\calS$ an {\it objective molecular structure}. The case $M=1$ reduces to (\ref{oas}).   Structures satisfying this definition are
not always associated with collections of actual molecules.  Non ``molecular'' examples include typical examples of 
ordered alloys,  nanotubes and fullerines.  Also, in any realistic example,  the atom described by $(i, \ell)$
should be the same species as atom $(k, \ell)$.

The value of these definitions rests on the empirical observation that collections of molecules are found to satisfy 
these rules.  The definition is also consistent with the construction of piecewise rigid origami.  In this case we consider a
collection of $N$ identical tiles $\calT_i = \bfc_i+ \calT, \ \  \bfc_i \in \R^2,\ i= 1, \dots, N$ with $\bfc_1 = \mathbf{0}$.  Normally, these are defined by
a crease pattern, so the $\calT_i$ are disjoint and  $\calR = \cup \overline{\calT}_i$ is a simply connected planar domain.  Suppose that each is deformed 
by a mapping $\bfy_i: \calT_i \to \R^3$, and consider the structure defined by $\bfy(\bfx) = \bfy_i(\bfx), \ \bfx \in \calT_i$.  
Then the origami  structure analogous of an objective molecular structure is the set of
deformations $\bfy_1, \dots, \bfy_N$, defined as above, such that for each $\bfx \in \calT$ and each $\bfz \in \calR$
there is an orthogonal tensor $\bfQ_i$ depending on $\bfx$ such that 
\beq
\bfQ_i(\bfx) (\bfy (\bfz)  - \bfy_1(\bfx)) + \bfy_i(\bfx+ \bfc_i) \in \bfy(\calR). \label{oos}
\eeq

For typical origami structures we would also impose the continuity and invertibility (if possible)  of $\bfy$.  Also, for classic origami 
$\bfy$ is piecewise rigid, but this need not be the case.  

The definitions (\ref{oas})-(\ref{oos}) are not so convenient for the design of structures or molecules.  
In fact, they imply a more useful underlying group structure.  We first observe that real atomistic structures are discrete,
and we add this to the definition of an objective structure: the structure contains no accumulation points.  We consider an
objective molecular structure $\calS$ defined by (\ref{oms}).  We consider isometries, written in conventional notation $(\bfQ|
\bfc)$, $\bfQ \in$ O(3), and $\bfc \in \R^3$.  Next we define the {\it isometry group of} $\calS$ as the set of all $(\bfQ|\bfc)$
such that
\beq
(\bfQ|\bfc)(\bfx_{k,\ell}) := \bfQ \bfx_{k,\ell} + \bfc = \bfx_{\Pi(k. \ell)}, \quad k = 1, \dots, N, \ \ \ell = 1, \dots, M,  \label{defiso}
\eeq
where $\Pi( \cdot, \cdot)$ is a permutation on two indices that preserves species in the sense given above. The natural
group product associated to this definition is composition of mappings 
\beq
(\bfQ_1| \bfc_1) (\bfQ_2 | \bfc_2) = (\bfQ_1 \bfQ_2 | \bfc_1 +  \bfQ_1 \bfc_2)   \label{gp}
\eeq
with the identity being $(\bfI | \mathbf{0})$.  Using these definitions, let $\calG$ be the isometry group of $\calS$.  

We claim that  $\calS$ is the orbit of Molecule 1, $\calM_1 = \{\bfx_{1,\ell}: \ell = 1, \dots M \}$,  under $\calG$.  To see
this, rearrange the definition of an objective molecular structure to read 
$ \bfQ_{i,j}\bfx_{p,q} + \bfx_{i,j} - \bfQ_{i,j}\bfx_{1,j} = \bfx_{\Pi(p,q)}$.  Here, to simplify the notation, we have suppressed the 
parametric dependence of the permutation $\Pi$ on $i,j$.  Thus, $g_{(i,j)} := (\bfQ_{i,j} | \ \bfx_{i,j}  -\bfQ_{i,j} \bfx_{1,j} )$
belongs to the isometry group $\calG$ for each $i=1, \dots N,\ j = 1, \dots, M$. But, $g_{(i,j)}$ operating on the $j^{th}$ atom
of Molecule 1 is, trivially,  $g_{(i,j)}(\bfx_{1,j}) = \bfx_{i,j}$.  So, the orbit of $\calM_1$ under $\calG$ is contained in $\calS$.
But $\calS$ contains the orbit of $\calM_1$ under $\calG$ by the definition (\ref{defiso}) of an isometry group.

This simple proposition obscures two facts.  First, it allows for molecules to be overlapping.  Once recognized, this is in fact a good
feature in terms of applications.  An example is the ethane molecule, C$_2$H$_6$, which, in terms of the present discussion,
can be considered as the orbit of C-H under its isometry group.  But, clearly, various elements of this group map the C of C-H to
itself.  It would not be useful to exclude these elements.  The second issue is discreteness.  To be realistic, the atomic structure
should be discrete.  Also, discreteness is a powerful hypothesis used extensively  in the known derivation of the discrete groups of isometries
presented, for example, in the International Tables of Crystallography.   

So, the question arises: could one have a non-discrete group of isometries $\calG$ and a molecule $\calM_1$ such that the orbit
of $\calM_1$ under $\calG$ is a discrete structure (and therefore realistic)?   To  show that this possibility is uninteresting, it is
sufficient to consider an objective atomic structure.

\begin{proposition}  Suppose $\calS$ is a discrete structure
which is the orbit of a nondiscrete isometry group $G$ applied to
a point $\bfx_1 \in \R^3$.  Then $\calS$ is a single point, a
pair of points, a periodic line of points $\{ i \bfe + \bfc,
\, i \in \Z \}$ in a direction
$\bfe$, or the union of two periodic lines of points
with the same period and contained on the same line:
$\{ i \bfe + \bfc, \, i \in \Z \} \cup
\{ (i + \lambda) \bfe + \bfc, \, i \in \Z \}$, $\lambda \ne 0$.
\label{prop1}
\end{proposition}

\noindent A proof is given in the Appendix. 

One should not conclude from this proposition that  non-discrete groups are not interesting!  In fact, it is a
main purpose of this paper to highlight their usefulness (Section  \ref{sect5}).

These results underlie extremely simple methods of constructing objective molecular structures
we call the {\it group theory method}.  Numerous examples are given below.
For atomic structures we simply assign atomic positions and species in, say, Molecule 1, and we take 
its orbit under a discrete group of isometries to generate a molecular structure.  In addition to the empirical 
observation of the widespread appearance of such structures, there are obvious theorems of stability.  
Since each atom of an objective atomic structure sees the same environment, then, 
for typical (i.e., frame-indifferent) descriptions of atomic forces, if one atom of the structure is in 
equilibrium, then all atoms are in equilibrium.  Similar arguments apply to stability \cite{james2006objective}. A recent thesis
\cite{steinbach} exploits this underlying structure for linear stability analysis in which many atoms are perturbed.

The group theory method applies also to origami structures.  In the simplest case, we consider a set of partly folded tiles.
For definiteness we can consider the partly folded structure $\calU$ of Figure \ref{fig:helical_origami}(a) bounded by the
four line segments $\overline{\bfy_1\bfy_2}, \overline{\bfy_2\bfy_3}, \overline{\bfy_3\bfy_4}, \overline{\bfy_4\bfy_1}$.
Now choose two isometries $g_1 = (\bfR_1|\bfc_1)$ and $g_2 = (\bfR_2|\bfc_2$) so that 
$g_1(\overline{\bfy_1\bfy_2}) = \overline{\bfy_4\bfy_3}$ and $g_2(\overline{\bfy_2\bfy_3}) = \overline{\bfy_1\bfy_4}$,
and arrange that $g_1$ and $g_2$ commute.  Then $\calG = \{ g_1^i g_2^j : \ i,j \in \Z \}$ is a group.  Now apply successively the
$\calG$ to all of $\calU$, not just its boundary.  The remarkable connection between Abelian groups and compatibility
means that the structure of all these images of $\calU$ fit together perfectly with no gaps.  Examples are shown in the various 
subfigures of  Figure \ref{fig:helical_origami}.  Since there are a lot of Abelian groups of isometries, and a lot of unit cells,
the method has broad scope for designing origami structures.  We look at the method in more detail in Section \ref{sect3.2}.






\section{1-D materials (nanotubes), helical origami}
\label{sect3}
The ubiquitous nanotube-like atomic structures, for example, carbon nanotubes, nanotubes BCN, GaN, and MoS$_2$, are generically helical structures. As a class of objective structures, helical structures are generated by applying the helical groups to an atom or a set of atoms in space. Two different helical structures can form geometrically compatible interfaces separating two phases. The concept of geometrical compatibility has been widely and successfully used to analyze hysteresis, fatigue, and reversibility in martensitic phase transformations \cite{bhattacharya_microstructure_2003, song_enhanced_2013}. Transforming one phase to the other by moving the phase boundary, the structure exhibits macroscopic twist and extension. Analogous ideas apply to designing helical Miura origami and its actuation.
\subsection{Helical groups and helical structures}
\begin{figure}[ht]
	\centering
	\includegraphics[width=\textwidth]{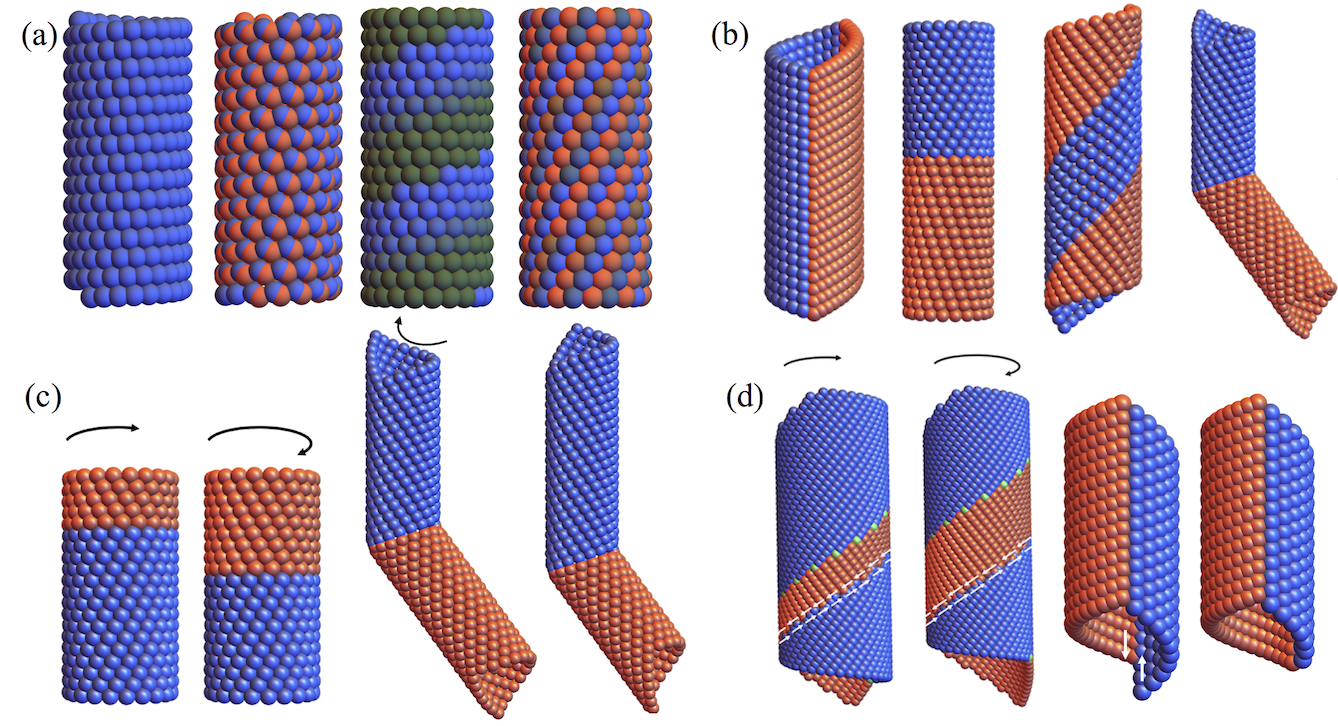}
	\caption{(a) Four types of helical groups. Each picture is the orbit of a single ball under the corresponding group and the coloring is according to the powers $s$ or $n$. (b) Four types of compatible interfaces between two helical structures. From left to right: vertical, horizontal, helical, and elliptical interfaces. (c) The horizontal and elliptical interfaces can move by transforming one phase to the other with no slips. The transformation induces macroscopic twist and extension. (d) The helical and vertical interfaces are rigid. Moving the interfaces by phase transformation will introduce slips (indicated by white arrows) on the other interfaces.}
	\label{fig:helical_structure}
\end{figure}
Helical groups are by definition discrete groups of isometries that contain no pure translations and do not fix a point in $\mathbb{R}^3$. Following the definition, a helical group is given by one of the four formulas \cite{feng2019phase}
\beqs 
& & \{ h^m  : m \in \Z \},   \label{G1} \\
& &  \{ h^m f^{s} : m \in \Z,\, s = 1,2  \},   \label{G2}  \\
& & \{ h^m g^n : m \in \Z,\, n = 1, \dots, i  \},    \label{G3}  \\
& &  \{ h^m g^n f^{s} : m \in \Z,\, n = 1, \dots, i,\, s = 1,2  \},  \label{G4} 
\eeqs
where
\begin{enumerate}
	\item  $h = (\bfQ_{\theta}| \tau \bfe + (\bfI - \bfQ_{\theta})\bfz \}$,
	$\bfQ_{\theta} \bfe = \bfe,\, |\bfe| =1,\, \bfz \in \mathbb R^3,\, \tau \in \mathbb{R}\setminus \{ 0\}$,
	is a screw displacement with an  angle $\theta$ that is an irrational multiple of $2 \pi$.
	\item  $g = (\bfQ_{\alpha}| (\bfI-\bfQ_{\alpha}) \bfz )$,
	$\bfQ_{\alpha} \bfe = \bfe$,
	is a proper rotation with  angle $\alpha = 2\pi/i,\, i \in \mathbb{N}, \, i \ne 0$.
	\item $f = (\bfQ|\,
	(\bfI-\bfQ) \bfz_1),\, \bfQ =-\bfI + 2 \bfe_1 \otimes \bfe_1,
	\, |\bfe_1| = 1,
	\bfe \cdot \bfe_1 = 0$ is a 180$^\circ$ rotation with
	axis perpendicular to $\bfe$.   Here, $\bfz_1 = \bfz  + \xi \bfe$,
	for some $\xi \in \R$.
\end{enumerate}
Among the four groups, (\ref{G1}) and (\ref{G3}) are Abelian, while (\ref{G2}) and (\ref{G4}) are not, because  $f$ does not commute with the
other elements.
Figure \ref{fig:helical_structure}(a) illustrates the four types of helical groups (\ref{G1})-(\ref{G4}), by applying the elements of the groups to a single atom position. The coloring is according to the powers $s$ or $n$.

Helical atomic or molecular structures are generated by applying the helical groups to an atom position or a set of positions in $\mathbb{R}^3$.  The structural parameters of the resulting helical structures are determined by the parameters of helical groups and the positions of atoms to which
the groups are applied.
Under the standard parameterization above, the nearest atomic points do not correspond to the nearest powers of generators. 
Thus, powers of generators are not good representatives of metric properties.  This causes difficulties in studying several typical problems in helical structures, e.g., compatible interfaces, phase transformations, etc. Therefore, a new parameterization of the groups is needed.  For definiteness, we consider the largest Abelian helical group (\ref{G3}).
Fortunately, under the standing assumption of non-degeneracy, (\ref{G3}) can be systematically reparameterized by its two nearest neighbor generators $g_1$ and $g_2$ having the forms 
\beqs
g_1 &=& (\bfQ_{\psi}| (\bfI - \bfQ_{\psi})\bfz + m_1 \tau \bfe), \nonumber \\
g_2 &=& (\bfQ_{\beta}| (\bfI - \bfQ_{\beta})\bfz + m_2 \tau \bfe), \label{defg1g2}
\eeqs
given by a rigorous algorithm in \cite{feng2019phase} \footnote{The reparameterization also applies to some rod groups that contain translations. }.
Choosing the appropriate domain of powers of generators, $g_1$ and $g_2$ generate exactly the same atom positions as (\ref{G3}); that is, the orbit of a point $\mathbf{x} \in \mathbb{R}^3$ under
\beq
\calG=\{g_1^p g_2^q : p \in \mathbb{Z}, q = 1,2,\dots, q^\star\}
\label{eq:helical_rep}
\eeq
produces the same structure as the original parameterization. 
(A formula for $q^\star$ can also be given, see \cite{feng2019phase}.) The reparameterization ensures that the nearest neighbors in powers $(p, q)$ correspond to the nearest neighbors in atomic positions. 
We employ the reparameterized helical group (\ref{eq:helical_rep}) and the concept of rank-1 compatibility (which is familiar in the study of martensitic phase transformations \cite{bhattacharya_microstructure_2003}) to study the compatible interfaces between two different helical structures. Specifically, the deformation from the domain of powers of generators to the two different helical structures induced by the group action is 
\beq
\bfy_i (p,q)= g_{1i}^p g_{2i}^q (\bfp_i) = \bfQ^i_{p \psi_i + q \beta_i} (\bfp_i - \bfz_i) + (p m_1^i + q m_2^i) \tau_i \bfe_i + \bfz_i, \label{eq:deformation} 
\eeq
where $i \in \{a, b\}$ indicates the parameters of phase $a$ or phase $b$.  The structural parameters $\{\psi_i, \beta_i, \bfp_i, \bfz_i, m_1^i, m_2^i, \tau_i, \bfe_i, \bfz_i\}$ determine the structures of the two phases. 

This formula (\ref{eq:deformation}) gives discrete atomic positions, but actually makes perfect sense if $p,q$ are real numbers.
Thus (\ref{eq:deformation}) gives an excellent smooth, nonoscillating interpolation of atomic positions.  Then, compatibility of helical phases
can be defined via the compatibility condition of continuum mechanics, i.e., interfaces are compatible if and only if the $(p,q)$ gradients are rank-1 connected. That is,
\beq
\nabla_{p, q} \bfy_a(\hat{p}(s), \hat{q}(s))  - \nabla_{p, q} \bfy_b(\hat{p}(s), \hat{q}(s)) =\bfa(s) \otimes \bfn(s),  \label{eq:rank1}
\eeq
where $(\hat{p}(s), \hat{q}(s))$ is the continuous interface on the reference domain, $\bfn(s) = (-\hat{q}'(s), \hat{p}'(s))$, and $s$ is the arc-length parameter.

In  \cite{feng2019phase} we characterized the four and only four types of compatible interfaces by finding the structural parameters and interfaces $(\hat{p}(s), \hat{q}(s))$ that satisfy (\ref{eq:rank1}). Examples of the compatible interfaces are shown in Fig. \ref{fig:helical_structure}(b): vertical, horizontal, helical, and elliptical interfaces. Among them, the horizontal and elliptical interfaces are mobile (Figure \ref{fig:helical_structure}(c)), whereas the vertical and helical interfaces are stabilized by the global compatibility of the structure (Figure \ref{fig:helical_structure}(d)). The phase transformation will induce macroscopic twist and extension for the horizontal and elliptical interfaces, while slip is required (and can be quantified) for the vertical and helical interfaces.

\subsection{Helical Miura origami}
\label{sect3.2}
\begin{figure}[ht]
	\centering
	\includegraphics[width=\textwidth]{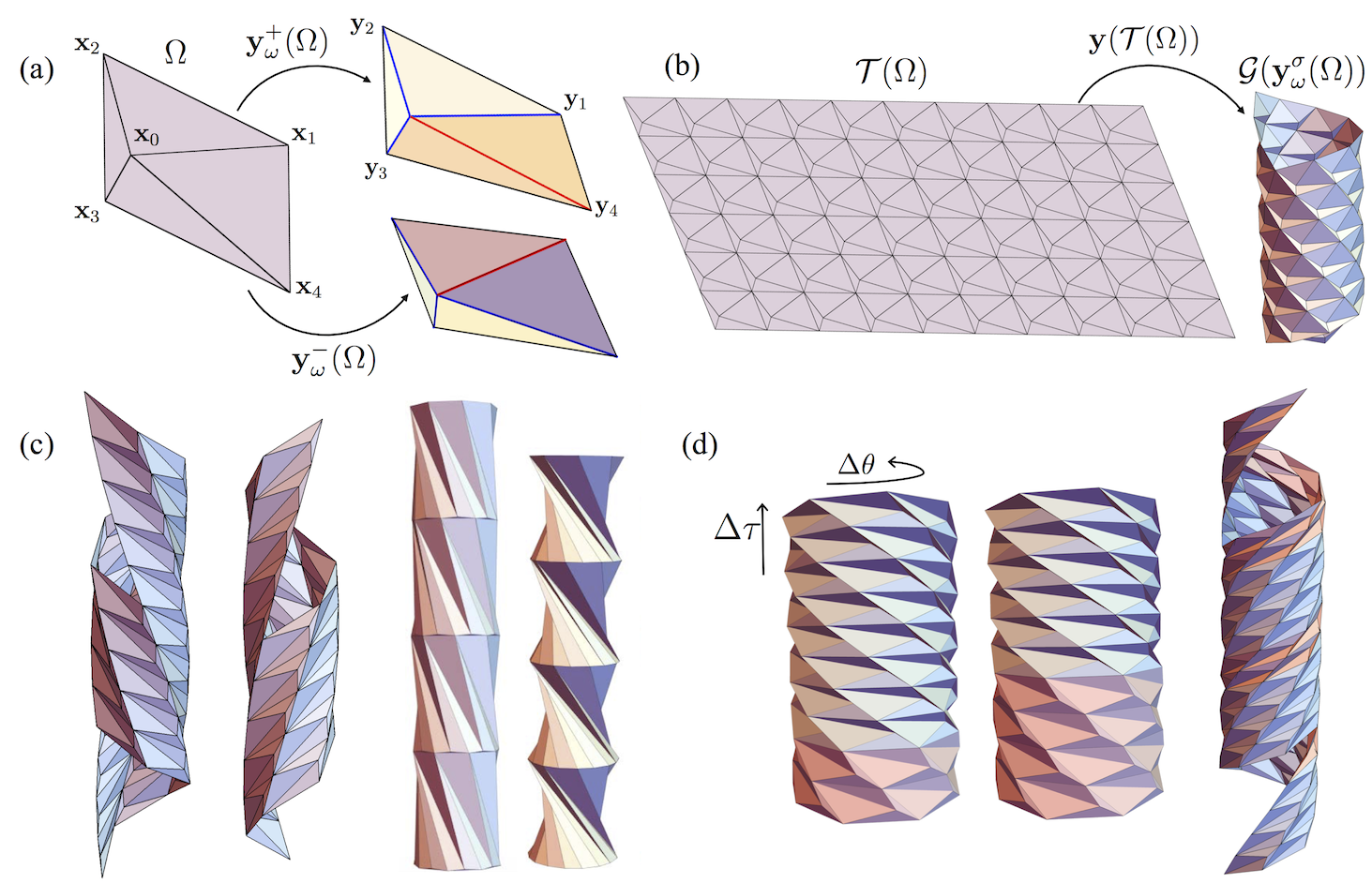}
	\caption{(a) The reference Miura parallelogram $\Omega$ and its partially folded states $\bfy_{\omega}^{\sigma}(\Omega)$. The folding kinematics $\bfy_{\omega}^{\sigma}$ have two choices of folding branch $\sigma \in \{\pm\}$ relating to different mountain-valley assignments. The blue/red lines indicate the mountains/valleys. (b) The reference tiling $\calT(\Omega)$ is rolled up to the HMO tiling $\calG(\bfy_{\omega}^{\sigma}(\Omega))$ by the deformation $\bfy$.
		(c) Examples of helical Miura origami with different chiralities and folding angles. (d) Horizontal and helical interfaces in helical Miura origami. The phase transformation from one phase to the other can induce twist $\Delta \theta$ and extension $\Delta \tau$.}
	\label{fig:helical_origami}
\end{figure}
Helical Miura origami (HMO) \cite{feng2020helical} is a cylindrical origami constructed by applying the helical or rod group to a partially folded unit cell
using the group theory method (Section \ref{sect1}). The  unit cell we choose is a partially folded Miura parallelogram $\Omega$ (Figure \ref{fig:helical_origami}(a)) with a four-fold vertex satisfying Kawasaki's condition, i.e., the opposite sector angles $\angle\bfx_1 \bfx_0 \bfx_2$ and $\angle \bfx_4 \bfx_0 \bfx_3$ sum to $\pi$. Up to overall isometry, the folding kinematics has one degree of freedom, the reference folding angle $\omega$, and two folding branches indicated by a topography parameter $\sigma \in \{\pm\}$ representing the so-called mountain-valley assignments.  The resulting deformations have been explicitly characterized by different approaches in \cite{feng2020helical, huffman1976curvature}. For our purposes the partially folded state of the Miura parallelogram is given by a deformation $\bfy_{\omega}^{\sigma} (\Omega)$
with positions of vertices $\bfy_i = \bfy_{\omega}^{\sigma} (\bfx_i)$, $i = 1, 2, 3, 4$. Here the function $\bfy_{\omega}^{\sigma}: \Omega \rightarrow \mathbb{R}^3$ is explicit and describes the deformation from the flat state $\Omega$ to the partially folded state with the reference folding angle $\omega$ and folding branch $\sigma$. 

We construct the HMO by taking the group action of $\calG=\{g_1^p g_2^q: p, q\in \mathbb{Z}\}$ on $\bfy_{\omega}^{\sigma}(\Omega)$ with the generators 
\beq
g_i = (\bfR_{\theta_i}|(\bfI - \bfR_{\theta_i}) \bfz + \tau_i \bfe), \quad i=1,2,	\label{eq:generator}
\eeq
in which $\bfR_{\theta_i} \in$ SO(3), $\theta_i \in (-\pi, \pi]$, $\tau_i \in \mathbb{R}$, $\bfz \in \mathbb{R}^3$, $\bfe\in \mathbb{R}^3$, $|\bfe|=1$, and $\bfz \cdot \bfe =0$ characterizing the rotation, rotation angle, translation, origin of the isometry, and rotation axis, respectively. These parameters are subject to a discreteness condition, 
\beqs
p^\star \theta_1 + q^\star \theta_2 &=& 2\pi, \nonumber \\
p^\star \tau_1 + q^\star \tau_2 &=&0, \label{eq:global}
\eeqs
for some integers $p^\star, q^\star \in \mathbb{Z}$. This condition is necessary and sufficient for the discreteness of $\calG$ (see Section \ref{sect5})
and is related to the absence of a ``seam'' when the cylindrical structure is formed by isometrically rolling up a periodic sheet of atoms (Figure \ref{fig:helical_origami}(b)).  (For an illustration of what happens when (\ref{eq:global}) fails, see Figure \ref{fig:nondiscrete_helical}(b)). The pair of integers 
$(p^\star, q^\star)$ is called the chirality.

According to the group theory method (Section \ref{sect1}), the generators $g_1$ and $g_2$ have only to obey the local compatibility of the edges of the adjacent unit cells $\bfy_{\omega}^{\sigma}(\Omega)$, $g_1(\bfy_{\omega}^{\sigma}(\Omega))$ and $g_2(\bfy_{\omega}^{\sigma}(\Omega))$.   Specifically, since isometries are affine, we need only satisfy
\beq
g_1(\bfy_4) = \bfy_1, ~g_1(\bfy_3) = \bfy_2,~ g_2(\bfy_1) = \bfy_2, ~g_2(\bfy_4) = \bfy_3. \label{eq:local}
\eeq

The commutativity of $g_1$ and $g_2$, i.e. $g_1g_2 = g_2 g_1$, ensures the compatibility of the fourth unit cell $g_1 g_2(\bfy_{\omega}^{\sigma}(\Omega))=g_2 g_1(\bfy_{\omega}^{\sigma}(\Omega))$, and all cells formed using higher powers of $g_1$ and $g_2$. By solving  (\ref{eq:global}) and (\ref{eq:local}) for fixed reference unit cell, $(p^\star, q^\star)$ and $\sigma$, one can find $0-4$ solutions for $\omega$ according to the numerical results in \cite{feng2020helical}. Such solutions correspond to compatible HMO structures. Some examples are presented in Figure \ref{fig:helical_origami}(c) with different chiralities $(p^\star, q^\star)$ and folding angles $\omega$. On the other hand, the construction is equivalent to a ``rolling-up" deformation $\bfy$ (referred to above) from a reference tiling $\calT (\Omega)$ to the HMO tiling $\calG(\bfy_{\omega}^{\sigma}(\Omega))$ (Figure \ref{fig:helical_origami}(b)), where $\calT = \{t_1^p t_2^q | (p,q)\in\mathbb{Z}^2\}$ is a translation group with generators $t_1=(\bfI|\bfx_1 - \bfx_4)$ and $t_2=(\bfI|\bfx_2-\bfx_1)$. Now we use an idea in \cite{ganor2016zig} to link the group of the reference domain to the group of the deformed domain and define an explicit form of the deformation $\bfy$. To this end, we first notice that the local compatibility condition (\ref{eq:local}) implies the compatibility of the folding kinematics $\bfy_{\omega}^{\sigma}(\bfx)$ as 
	\beqs
	\bfy_{\omega}^{\sigma}(\bfx) &&= g_1^{k_1} g_2^{k_2} (\bfy_{\omega}^{\sigma} (t_1^{-k_1} t_2^{-k_2} (\bfx))) \nonumber \\
	&&= \bfR_{k_1 \theta_1 + k_2 \theta_2} \bfy_{\omega}^{\sigma}(\bfx-k_1(\bfx_1-\bfx_4) - k_2 (\bfx_2 - \bfx_1)) \nonumber \\
	&& +(\bfI - \bfR_{k_1 \theta_1 + k_2 \theta_2}) \bfz + (k_1 \tau_1 + k_2 \tau_2)\bfe,
	\quad \bfx \in \calI_{k_1 k_2}, \label{eq:localgroup}
	\eeqs
	where $\calI_{k_1 k_2} = t_1^{k_1} t_2^{k_2} (\Omega) \cap \Omega $, for $k_1, k_2 \in \{0,1\}$. Clearly, the set $\cup \calI_{k_1 k_2}$ contains two adjacent edges of the unit cell and Equation (\ref{eq:localgroup}) ensures that the four adjacent unit cells are compatible. Then we extend the reference domain to $\calT(\Omega)=t_1^p t_2^q(\Omega)$ and the deformation is extended to
	\beqs
	\bfy(\bfx)  &&= g_1^p g_2^q (\bfy_{\omega}^{\sigma}(t_1^{-p} t_2^{-q} (\bfx))) \nonumber \\
	&&= \bfR_{p \theta_1 + q \theta_2} \bfy_{\omega}^{\sigma}(\bfx-p (\bfx_1-\bfx_4) - q (\bfx_2 - \bfx_1)) \nonumber \\
	&& +(\bfI - \bfR_{ p \theta_1 + q\theta_2}) \bfz + (p \tau_1 + q \tau_2)\bfe,
	\quad \bfx \in t_1^p t_2^q(\Omega), 
	\label{eq:globalgroup}
	\eeqs
	where $(p,q)\in\mathbb{Z}^2$. One can easily show that, by (\ref{eq:global}) and (\ref{eq:localgroup}), the edges in $\calT(\Omega)$ deformed by $\bfy(\bfx)$ are all compatible, and therefore, the resulting HMO is compatible. 

The existence of multiple solutions implies that HMO is multistable for an appropriate unit cell $\Omega$, folding branch $\sigma$, and chirality $(p^\star, q^\star)$. These different solutions can be treated as different ``phases" in the scope of phase transformation. Following the generalized local and global compatibilities (see \cite{feng2020helical}), an HMO can have multiple phases separated by compatible interfaces and still remain compatible as a cylindrical structure (Figure \ref{fig:helical_origami}(d)). Different phases have different folding angles or folding branches, and therefore generally they have different structural parameters.
Inspired by the atomic phase transformation, we are able to transform one phase to the other through compatible interfaces and induce overall twist and extension. This mechanism is applicable for designing origami actuators, artificial muscles, and robotics.

\section{2-D materials, 2-D origami}
\label{sect4}

Since the discovery of superconductivity in twisted bilayer graphene \cite{cao}, there has been a resurgence of
interest in 2-D structures, especially with particular Moir\'e patterns \cite{yoo}.   Origami design, on the other hand, suggests 
ways of designing nanostructures with particular patterns of neighbors.

\subsection{A family of 2D origami structures with degeneracy}

Degeneracies in origami design, i.e., the many ways to fold a crease pattern, are particularly interesting in the context of the search for novel nanostructures.
For example, if we identify the vertices of an origami structure with atomic positions, degeneracy gives us many structures with
the same nearest neighbor distances for all the atoms.  This follows simply from the fact that an origami deformation is piecewise isometric.

Below, we discuss degeneracies in the context of a simple, yet fascinating, family of origami: \textit{rigidly and flat-foldable quadrilateral mesh origami} \cite{huffman1976curvature,miura1985method,hull1994mathematics,tachi2009generalization,lang2011origami,lang2018,designs}. Despite being a well-studied family of origami over the years, interest in their degeneracies is a recent development \cite{hull2014counting,lang2018,dieleman2020jigsaw} with many intriguing directions for further exploration.  Here, we show that there are tessellations in this family that can be folded a huge number of ways.

\subsubsection{On quad-meshes that can be rigidly folded flat}

\begin{figure}[t!]
\centering
\includegraphics[width =\textwidth]{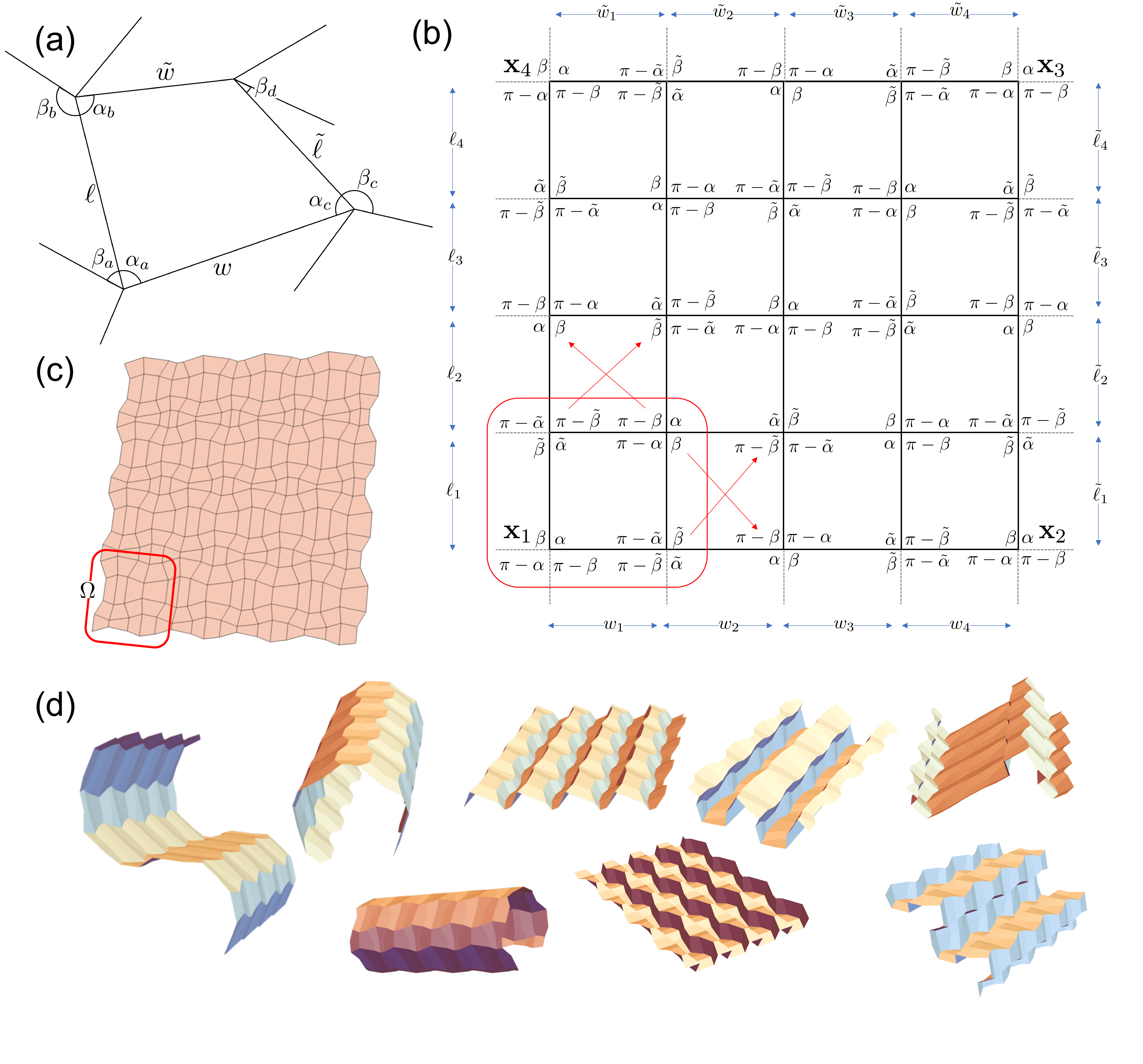}
\vspace{-10mm}
\caption{(a) Notation for the crease pattern surrounding a single tile (sector angles and lengths) for which opposite sector angles sum to $\pi$.   (b) Description of a highly degenerate unit cell.  The checkerboard schematic indicates the topology only, i.e., the angles are not right angles. (c) A tessellation emerging from this analysis and (d) a few examples of the 65534 ways this tessellation can be folded by varying only the mountain-valley assignments. }
\label{fig:Geometry}
\end{figure}

In \cite{designs} we give necessary and sufficient conditions for the flat foldability of a piecewise rigid quadrilateral 
mesh sheet such as that shown in Figure \ref{fig:Geometry}(c).  The conditions are formulated in terms of an efficient algorithm -- which (incidentally) can be used to design a myriad of deployable structures with origami \cite{inverse}. 

On the topic of degeneracies, we build on ideas from \cite{designs}: As derived there, the question of whether or not a flat crease pattern, like the one shown in Figure \ref{fig:Geometry}(a), is rigidly and flat-foldable can be addressed succinctly in terms of products of so-called \textit{fold angle multipliers}.  Fold angle multipliers are the functions
\begin{equation}
\begin{aligned}\label{eq:foldAngMult}
\mu_2(\alpha, \beta, \sigma) :=  \frac{-\sigma  + \cos \alpha \cos \beta + \sin \alpha \sin \beta}{ \cos \beta - \sigma \cos \alpha}, \quad \mu_1 ( \alpha, \beta,- \sigma) := \mu_2(\alpha, \pi - \beta, -\sigma) 
\end{aligned}
\end{equation}  
defined for sector angles $\alpha, \beta \in (0, \pi)$, $(\alpha, \beta) \neq (\pi/2,\pi/2)$ and mountain-valley assignment $\sigma \in \mathcal{MV}(\alpha, \beta)$ indicated by
\begin{equation}
\begin{aligned}\label{eq:MVDef}
\mathcal{MV}(\alpha, \beta) :=  \left\{ \begin{array}{l} -1  \qquad \text{ if } \alpha = \beta \neq \pi/2 \\ +1   \qquad \text{ if } \alpha = \pi - \beta \neq \pi/2  \\ \pm 1  \qquad  \text{ if }  \alpha \neq \beta \neq \pi - \beta \end{array} \right\} .
\end{aligned}
\end{equation}
The crease pattern Figure \ref{fig:Geometry}(a) is parameterized by seven sector angles 
\begin{equation}
\begin{aligned}
&\alpha_a, \beta_a, \alpha_b, \beta_b, \alpha_c, \beta_c, \beta_d \in (0,\pi), \quad  \alpha_d := 2 \pi - \alpha_a - \alpha_b - \alpha_c \in (0,\pi) \quad  
\end{aligned}
\end{equation}
and we also assume for simplicity a {\it right angle restriction}:
\begin{equation}
(\alpha_a, \beta_a),\  (\alpha_b, \beta_b), \ 
(\alpha_c, \beta_c),\  (\alpha_d, \beta_d) \neq (\pi/2, \pi/2).
\end{equation}
Taking these sector angles as given, the fold angle multipliers at each vertex satisfy
\begin{equation}
\begin{array}{lll}
\mu_{2a}(\sigma) := \mu_2(\alpha_a, \beta_a, \sigma), & \mu_{1a}(-\sigma) := \mu_1(\alpha_a, \beta_a, -\sigma), & \sigma \in \mathcal{MV}(\alpha_a, \beta_a),  \\
\mu_{2b}(\sigma) := \mu_2(\alpha_b, \beta_b,\sigma), & \mu_{1b}(-\sigma) := \mu_1(\alpha_b, \beta_b, -\sigma), & \sigma \in \mathcal{MV}(\alpha_b, \beta_b),  \\
\mu_{2c}(\sigma) := \mu_2(\alpha_c, \beta_c, \sigma), & \mu_{1c}(-\sigma) := \mu_1(\alpha_c, \beta_c,-\sigma), & \sigma \in \mathcal{MV}(\alpha_c, \beta_c),  \\
\mu_{2d}(\sigma) := \mu_2(\alpha_d, \beta_d, \sigma), & \mu_{1d}(-\sigma) := \mu_1(\alpha_d, \beta_d,-\sigma), & \sigma \in \mathcal{MV}(\alpha_d, \beta_d). 
\end{array}
\end{equation}
In this formalism, the crease pattern is rigidly and flat-foldable if and only if 
\begin{equation}
\begin{aligned}\label{eq:loop}
&\mu_{1c}(-\sigma_c) \mu_{2d}(\sigma_d)\mu_{1b}(-\sigma_b) \mu_{2a}(\sigma_a) = 1,  \quad \text{ for some } \quad (\sigma_a, \sigma_b, \sigma_c, \sigma_d) \in \mathcal{MV}_{abcd}
\end{aligned}
\end{equation}
where $\mathcal{MV}_{abcd}:= \mathcal{MV}(\alpha_a, \beta_a) \times \mathcal{MV}(\alpha_b, \beta_b) \times \mathcal{MV}(\alpha_c, \beta_c) \times \mathcal{MV}(\alpha_d, \beta_d)$.

A key point for revealing degeneracy is the ``for some" in the statement.  Notice that each vertex has generically two choices of signs (\ref{eq:MVDef}).  So there are (naively) up to $16$ distinct collections of signs with which to test whether loop condition (\ref{eq:loop}) holds for a given set of sector angles.  These signs represent different mountain-valley assignments. So if the loop condition holds for distinct choices of $(\sigma_a, \sigma_b, \sigma_c, \sigma_d), (\tilde{\sigma}_a, \tilde{\sigma}_b, \tilde{\sigma}_c, \tilde{\sigma}_d), \ldots \in \mathcal{MV}_{abcd}$, then the crease pattern can be folded from flat to fold-flat along the distinct mountain-valley assignments indicated by each such $(\sigma_a, \sigma_b, \sigma_c, \sigma_d), (\tilde{\sigma}_a, \tilde{\sigma}_b, \tilde{\sigma}_c, \tilde{\sigma}_d), \ldots$ A natural question to ask then is:
\begin{itemize}
\item What are the most degenerate rigidly and flat-foldable crease patterns surrounding a single tile? 
\end{itemize}
That is, what crease patterns give the greatest number of distinct mountain-valley assignments satisfying (\ref{eq:loop}).  Through a combined  analytical and numerical approach, it is possible to justify the following theorem.
\begin{theorem}
The most degenerate families of rigidly and flat-foldable crease patterns surrounding a single tile can be folded along exactly six distinct mountain-valley assignments indicated by\footnote{Here and in the remainder of this section, we drop the ``1" when referencing mountain-valley assignments, i.e., the quantities belonging to the set (\ref{eq:MVDef}).} 
\begin{equation}
\begin{aligned}\label{eq:MVAssignements}
\left(\begin{array}{cc} \sigma_b & \sigma_d \\ \sigma_a  & \sigma_c  \end{array}\right) \in \Big\{ \left(\begin{array}{cc} + & + \\ + & + \end{array}\right), \left(\begin{array}{cc} - & - \\ - & - \end{array}\right), \left(\begin{array}{cc} + & + \\ - & - \end{array}\right),\left(\begin{array}{cc} - & - \\ + & + \end{array}\right), \left(\begin{array}{cc} + & - \\ + & - \end{array}\right) , \left(\begin{array}{cc} - & + \\ - & + \end{array}\right)\Big\} .
\end{aligned}
\end{equation} 
There are exactly three such families:
\begin{enumerate}
\item[(i)] $\alpha_c = \pi - \alpha_a, \beta_c = \pi -\beta_a, \alpha_d = \pi - \alpha_b, \beta_d = \pi - \beta_b$ and $\alpha_a, \beta_a, \alpha_b, \beta_b \in (0,\pi)$ satisfy
\begin{equation}
\begin{aligned}\label{eq:signRatio}
\frac{\sin \beta_b}{\sin \alpha_b} = \frac{\sin \beta_a}{\sin \alpha_a} \neq 1.
\end{aligned}
\end{equation}
\item[(ii)] Exchange the roles of $(\alpha_b, \beta_b)$ and $(\alpha_c, \beta_c)$ in (i);
\item[(iii)] $\alpha_d = \pi - \alpha_a, \alpha_c = \pi - \alpha_b, \beta_d = \beta_a,  \beta_c = \beta_b$ and  $\alpha_a, \beta_a, \alpha_b, \beta_b \in (0,\pi)$ satisfy (\ref{eq:signRatio}).
\end{enumerate}
\end{theorem}

\subsubsection{A highly foldable family of $4 \times 4$ tessellations}

Let us focus on the family of crease patterns in (iii) above.  In particular, consider four sector angles $\alpha, \tilde{\alpha}, \beta, \tilde{\beta} \in (0,\pi)$ such that 
\begin{equation}
\begin{aligned}
\frac{\sin \beta}{\sin \alpha} = \frac{\sin \tilde{\beta}}{\sin \tilde{\alpha}} \neq 1.
\end{aligned}
\end{equation}
Let us further consider an overall quad-mesh tessellation indicated topologically by the checkerboard in Figure \;\ref{fig:Geometry}(b).  To populate the sector angles on this quad-mesh, we first isolate the the lower left quad (in red) and, in the local notation of \ref{fig:Geometry}(a), we set $\alpha_a = \alpha$, $\beta_a = \beta$, $\alpha_b = \tilde{\alpha}$, $\beta_b = \tilde{\beta}$ and $\alpha_c,\beta_c, \alpha_d, \beta_d$ in (iii). This yields the description of the sector angles shown.  This panel in isolation can be folded along six distinct mountain-valley assignments.  We then move on to the adjacent panel (either above or to the right) and attempt to prescribe it so as to fold degenerately as in the family in (iii).  There is exactly one way to do this: The sector angles diagonal to each other are directly related by the rules in (iii). Since we know the sector angles around two of four vertices, we use this relationship to determine the other two vertices.  We then iterate using this basic fact.  This iteration leads to the sector angles displayed in the $4 \times 4$ checkerboard in the figure.  Notice the left boundary and right boundary have the same sector angles.  Similarly, the bottom and top boundary also have the same sector angles.  In other words, iteration produces a $4 \times 4$ mesh that is periodic in the sector angles. 

Can a mesh with these sector angles be tessellated? The crux of the matter is the side lengths. Using the notation of Figure \;\ref{fig:Geometry}(b), one needs $\ell_i = \tilde{\ell}_i$ and $w_i = \tilde{w}_i$.  These quantities, however, cannot be prescribed arbitrarily. Recalling again the notation in Figure \ref{fig:Geometry}(a), the side lengths are related to the interior sector angles of the quadrilateral by the transformation 
\begin{equation}
\begin{aligned}\label{eq:lengthsAndAngles}
\left(\begin{array}{c} \tilde{\ell} \\ \tilde{w} \end{array}\right)=  \left(\begin{array}{cc} \frac{-\sin \alpha_b}{\sin (\alpha_a + \alpha_b + \alpha_c) } &\frac{ \sin (\alpha_a + \alpha_b)}{\sin (\alpha_a + \alpha_b + \alpha_c) }\\ \frac{\sin (\alpha_a + \alpha_c)}{\sin (\alpha_a + \alpha_b + \alpha_c) } &\frac{-\sin \alpha_c}{\sin (\alpha_a + \alpha_b + \alpha_c) } \end{array}\right)\left(\begin{array}{c}  \ell \\ w \end{array}\right).
\end{aligned}
\end{equation}
Consequently, we prescribe the side lengths $\ell_1,\ldots, \ell_4$ and $w_1, \ldots, w_4$ and sector angles as shown in Figure \ref{fig:Geometry}(b); then every other side length of the crease pattern, including $\tilde{\ell}_1,\ldots, \tilde{\ell}_4$ and $\tilde{w}_1, \ldots, \tilde{w}_4$, is uniquely determined by iterating with the condition in (\ref{eq:lengthsAndAngles}).  Remarkably, we have the following identities for this procedure
\begin{equation}
\begin{aligned}\label{eq:ElliWi}
&\tilde{\ell}_i \equiv \tilde{\ell}_i(\alpha, \beta, \tilde{\alpha}, \tilde{\beta}, \ell_1,\ldots, \ell_4, w_1, \ldots, w_4) = \ell_i, \\
&\tilde{w}_i \equiv \tilde{w}_i(\alpha, \beta, \tilde{\alpha}, \tilde{\beta}, \ell_1,\ldots, \ell_4, w_1, \ldots, w_4) = w_i ,
\end{aligned}
\end{equation}
for all $i = 1,2,3,4$.  

We remark that there are choices of the parameters $\alpha, \beta, \tilde{\alpha}, \tilde{\beta}$, $\ell_1,\ldots, \ell_4$,$w_1, \ldots, w_4$ that produce unphysical side lengths on the interior of the mesh, i.e., lengths that evaluate to a non-positive number.  However, it is not difficult to find a family of parameters which  produces a physical $4 \times 4$ quad mesh crease pattern.  Let $\Omega \equiv \Omega(\alpha, \beta, \tilde{\alpha}, \tilde{\beta}, \ell_1,\ldots, \ell_4, w_1, \ldots, w_4)$ denote one such valid crease pattern, and let $\mathbf{x}_1,\mathbf{x}_2, \mathbf{x}_3, \mathbf{x}_4$ denote the ``four corner points" indicated in Figure \ref{fig:Geometry}(b).  Because of the identities in (\ref{eq:ElliWi}), we obtain a valid tessellation by taking the orbit of the unit cell $\Omega$ under the action of a translation group; explicitly,
\begin{equation}
\begin{aligned}\label{eq:tOmega}
\mathcal{T} \Omega = \big\{ t_1^p t_2^q (\Omega) \colon p, q \in \mathbb{Z}\big\}, \quad t_1 = (\mathbf{I} | \mathbf{x}_2 - \mathbf{x}_1), \quad t_2 = (\mathbf{I} | \mathbf{x}_4 - \mathbf{x}_1)
\end{aligned}
\end{equation}
parameterizes the tessellation.  One such example is provided in Figure \ref{fig:Geometry}(c).

These tessellations have the property that \textit{any} of their isolated $3 \times 3$ meshes can fold in the six ways indicated by the theorem.  We also know from \cite{designs} that a marching algorithm, prescribing the sector angles, side lengths and mountain-valley assignments on the left and bottom boundary of the pattern, completely determines the pattern and its kinematics.  Let us imagine we apply the sector angles and side lengths in the marching algorithm to be consistent with the tessellations given above.  The question then is: What collections of mountain-valley assignments will yield the tessellation (and, by extension, its kinematics along the prescribed mountain-valley assignment)? To answer this question, it is easiest to start simple and build. Consider the $3 \times 3$ lower-left corner of the tessellation. For the marching algorithm, apply the boundary sector angles and lengths consistent with the tessellation, and the mountain-valley assignments, for instance, as 
\begin{equation}
\begin{aligned}
\left(\begin{array}{ccc} + &  ?  & ?  \\ - & ? & ? \\ + & + & + \end{array}\right) .
\end{aligned}
\end{equation}
What will emerge? We can quickly convince ourselves using (\ref{eq:MVAssignements}) that the mountain-valley assignment that emerges from the algorithm is 
\begin{equation}
\begin{aligned}
\left(\begin{array}{ccc} + &  +  & +  \\ - & - & - \\ + & + & + \end{array}\right) 
\end{aligned}
\end{equation}
and the desired crease pattern is produced. Alternatively, if we alter the mountain-valley assignment as 
\begin{equation}
\begin{aligned}
\left(\begin{array}{ccc} - &  ?  & ?  \\ - & ? & ? \\ + & + & + \end{array}\right) , \quad \text{ then we get } \quad \left(\begin{array}{ccc} - &  -  & -  \\ - & - & - \\ + & + & + \end{array}\right)  ,
\end{aligned}
\end{equation}
yet the same crease pattern (consistent with the tessellation) is produced.  However, if we alter the mountain valley assignment as 
\begin{equation}
\begin{aligned}
\left(\begin{array}{ccc} - &  ?  & ?  \\ - & ? & ? \\ + & + & - \end{array}\right), \quad \text{ then we run into a problem: } \quad \left(\begin{array}{ccc} - &  -  & {\color{red}?}  \\ - & - &{\color{red} ?} \\ + & + & - \end{array}\right). 
\end{aligned}
\end{equation}
There is no consistent mountain-valley assignment in the listing  (\ref{eq:MVAssignements}).  Therefore, the algorithm cannot possibly produce the desired tessellation. Accounting for this dead end, there is a clear pattern to produce the tessellation by the marching algorithm: 
\begin{itemize}
\item Apply the mountain-valley assignments on either the left boundary or bottom boundary to be all the same (i.e., all $+,+,\ldots$ or all $-,-,\ldots$).
\item Apply the remaining mountain-valley assignments arbitrarily.  
\end{itemize}

A counting argument then furnishes the number of ways that these special crease patterns can be folded: if we consider a subset of the tessellation in (\ref{eq:tOmega}) with $M \times N$ interior vertices, then it can be folded along 
\begin{equation}
\begin{aligned}
2^M +2^N - 2 
\end{aligned}
\end{equation}
distinct mountain-valley assignments.  In  Figure \ref{fig:Geometry}(d), we provide eight of the 65,534 distinct ways of folding the crease pattern Figure \ref{fig:Geometry}(c).

\subsection{Objective non-isometric origami}
\begin{figure}[ht]
	\centering
	\includegraphics[width=\textwidth]{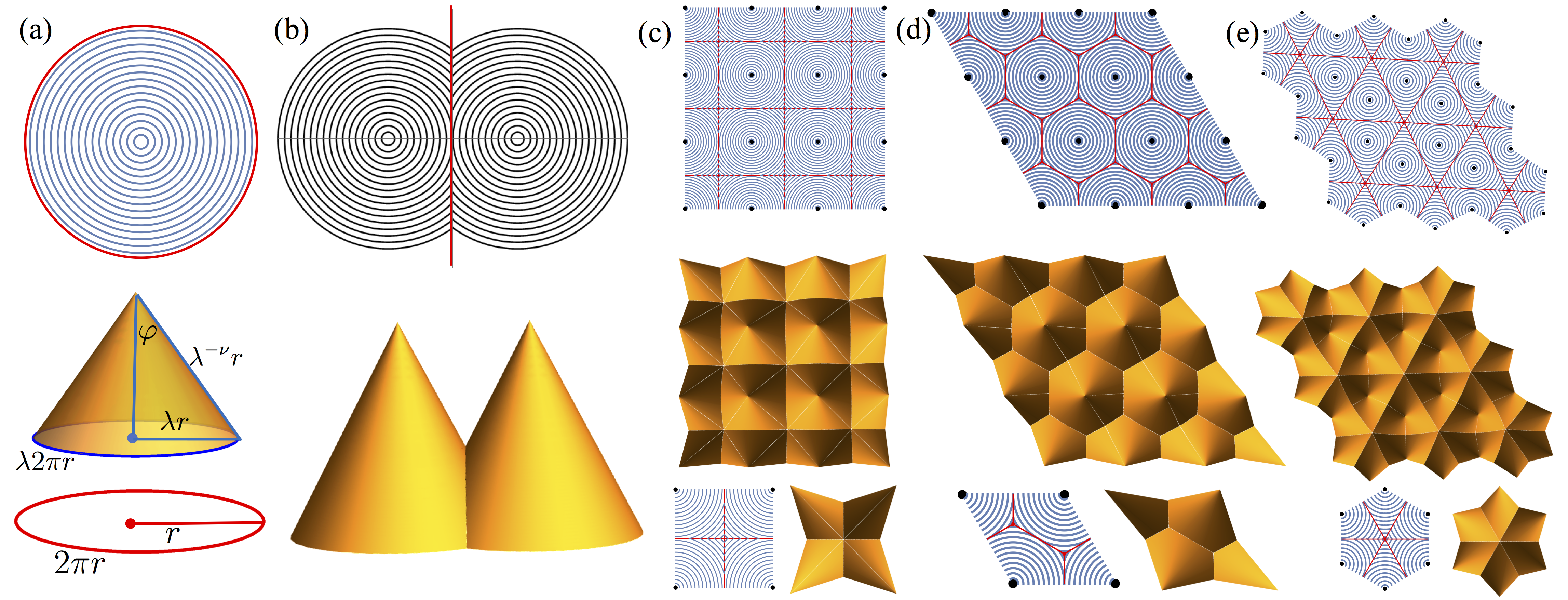}
	\caption{(a) A circular director pattern and its corresponding actuated cone. The circumference $2\pi r$ contracts by a factor $\lambda$ upon actuation. The in-material radius $r$ extends by $\lambda^{-\nu}$, since it is perpendicular to the director. The cone angle $\varphi$ is then given by $\varphi = \arcsin \lambda^{1+\nu}$. These facts are encoded in the cone deformation $\bfy_c(\bfr)$ in the text.
		(b) Two circular director patterns separated by a bisecting straight-line interface (red line). The actuated state is two cones meeting together with the same height. (c)-(e) 2D symmetric pattern and their actuated configurations. The centers of cones (black dots) form square, triangular, and hexagonal lattices.}
	\label{fig:cones}
\end{figure}
Unlike isometric origami, which is made of nearly unstretchable materials such as paper, \textit{non-isometric origami} is made of active materials carefully patterned into a sheet. The patterned sheet, in turn, responds to stimuli by origami deformations not isometric to the plane \cite{modes2011blueprinting, MWSPIE:12, plucinsky2016programming, Paul_non-isometric, warner2020topographic}.  Liquid crystal elastomers (LCEs) \cite{warner2007liquid} are active materials that can have significant length change along their ordering direction, the {\it director}, a unit vector $\bfn \in \R^3$. Driven by heat, light, or solvent, the 2D LCE sheet with programmed director field exhibits local spontaneous deformation described by the stretch tensor
\beq
\bfU_{\bfn} = \lambda \bfn \otimes \bfn + \lambda^{-\nu} \bfn^{\perp} \otimes \bfn^{\perp}, \label{eq:lcedeform}
\eeq
where $\bfn$ is the director and $\bfn^{\perp} \cdot \bfn = 0, |\bfn^{\perp} | = 1$. That is, the LCE sheet has a contraction $\lambda<1$ along the director and an elongation $\lambda^{-\nu}$ along $\bfn^{\perp}$ with the {\it optothermal Poisson ratio} $\nu$. Despite having the intrinsic metric change locally, it is still difficult to determine the macroscopic shape change of the entire pattern, but symmetry helps. Figure \ref{fig:cones}(a) describes the canonical example of shape-programming with LCEs \cite{modes2011gaussian,de2012engineering,ware2015voxelated}.  Top, in the figure, is a circular director pattern in which the director is parallel to the concentric circles.  The actuated state is a cone that respects the symmetry and the metric change. Specifically, the circumference $2 \pi r$ contracts by a factor $\lambda$ and the in-material radius $r$ extends by $\lambda^{\nu}$, since they are parallel or perpendicular to the director. Then the cone angle $\varphi$ is given by $\varphi = \arcsin \lambda^{1+\nu}$, as depicted in Figure \ref{fig:cones}(a).
	This induces the following cone deformation that encodes all the facts mentioned above and maps the reference domain to a cone:
	\beq
	\bfy_c(\bfx) = \lambda r (\bfe_r - \cot \varphi \bfe_3), \label{eq:cone_deformation}
	\eeq
	where $\bfx = r \bfe_r = r (\cos\theta \bfe_1 + \sin\theta \bfe_2)$ is the position on the circular pattern in which $(r,\theta)$ are the corresponding polar coordinates, and $\{\bfe_1, \bfe_2, \bfe_3 \}$ is the standard orthonormal basis for $\mathbb{R}^3$. 

As shown in \cite{feng2020evolving}, this basic design can be used as a building block for a large class of non-isometric origami: Two circular pattern with bisecting straight-line interface can form two equal-height cones with parallel axes after actuation, as depicted in Figure \ref{fig:cones}(b). Furthermore, symmetrically patterned circular patterns with bisecting interfaces can form {\it objective non-isometric origami} in Figures \ref{fig:cones}(c)-(e). We construct three examples by applying the 2D translation group on the ``unit cell". The unit cell $\Omega$ of the reference domain is a square, a rhombus, or a hexagon; see  Figure~\ref{fig:cones}, last of (c) , (d) and (e) respectively. The translation group, $\calT=\{t_1^p t_2^q: (p, q) \in \mathbb{Z}^2\}$ with $t_1=(\bfI |\bft_1)$ and $t_2 = (\bfI | \bft_2)$, generates the 2D tiling 
\beq
\mathcal{T} \Omega = \{ t_1^p t_2^q (\Omega)  \colon p,q \in \mathbb{Z}\}
\eeq
with translation symmetry.
The translation group $\hat{\calT}=\{\hat{t}_1^p \hat{t}_2^q: (p, q) \in \mathbb{Z}^2\}$ for the deformed domain is also two dimensional, but with linearly rescaled translations calculated from (\ref{eq:cone_deformation}). Specifically, the group generator $\hat{t}_i=(\bfI |\hat{\bft}_i)$ for the deformed domain has $\hat{\bft}_i = \lambda \bft_i$, for $i=1,2$.
We list the unit cells and translation groups in detail, for the examples in Figures \ref{fig:cones}(c)-(e):
\begin{enumerate}
	\item Figure \ref{fig:cones}(c). The four centers (black dots) in the unit cell are located at $\bfp_1={\bf 0}, \bfp_2 = \bfe_1, \bfp_3 = \bfe_1 + \bfe_2, \bfp_4=\bfe_2$. The generators for the reference domain are $t_1 = (\bfI | \bfe_1)$ and $t_2=(\bfI | \bfe_2)$. The generators for the deformed domain are $\hat{t}_1 = (\bfI| \lambda \bfe_1)$ and $\hat{t}_2= (\bfI | \lambda \bfe_2)$.
	
	\item  Figure \ref{fig:cones}(d). The four centers in the unit cell are located at $\bfp_1={\bf 0}, \bfp_2 = \bfe_1, \bfp_3 =1/2 \bfe_1 +\sqrt{3}/2 \bfe_2, \bfp_4=-1/2 \bfe_1 + \sqrt{3}/2 \bfe_2$. The generators for the reference domain are $t_1 = (\bfI | \bfe_1)$ and $t_2=(\bfI | -1/2 \bfe_1 + \sqrt{3}/2 \bfe_2)$. The generators for the deformed domain are $\hat{t}_1 = (\bfI | \lambda \bfe_1)$ and $\hat{t}_2= (\bfI | \lambda (-1/2 {\bfe}_1 + \sqrt{3}/2 {\bfe}_2))$.
	
	\item  Figure \ref{fig:cones}(e). The six centers in the unit cell are located at $\bfp_i = \bfR(\frac{i \pi}{6} )\bfe_1, i =1,2,\dots,6$, and $\bfR(.)$ is a rotation on $\bfe_1,\bfe_2$ plane. The generators for the reference domain are $t_1 = (\bfI | 3/2\bfe_1 + \sqrt{3}/2 \bfe_2)$ and $t_2=(\bfI | \sqrt{3} \bfe_2)$. The generators for the deformed domain are $\hat{t}_1 = ({\bfI} | \lambda(3/2{\bfe}_1 + \sqrt{3}/2 {\bfe}_2))$ and $\hat{t}_2= ({\bfI} | \lambda \sqrt{3}{\bfe}_2)$.
\end{enumerate}
Again, we follow exactly the same idea in Section \ref{sect3.2} and \cite{ganor2016zig} to explain the method of deriving an explicit deformation 
$\bfy(\bfx)$ that maps the reference tiling to the deformed tiling. To this end, we first assume the deformation that maps the reference unit cell to the deformed unit cell (see the last row of Figures \ref{fig:cones}(c)-(e)) is $\bfy_u(\Omega)$. This deformation can be derived by combining the cone deformations $\bfy_c$ for different subregions that belong to different director patterns while preserving compatibility at the boundaries of the cones. 
	The resulting deformation $\bfy_u$ is given by
	\beq
	\bfy_u(\bfx) = \hat{t}_1^{k_1} \hat{t}_2^{k_2}(\bfy_u(t_1^{-k_1} t_2^{-k_2}(\bfx))), \quad \bfx \in t_1^{k_1} t_2^{k_2}(\Omega) \cap \Omega,
	\eeq
	for $k_1, k_2 \in \{0, 1\}$. Finally the deformation $\bfy(\bfx)$ for the extended reference domain $t_1^p t_2^q(\Omega)$ is 
	\beq
	\bfy(\bfx) = \hat{t}_1^{p} \hat{t}_2^{q}(\bfy_u(t_1^{-p} t_2^{-q}(\bfx))), \quad \bfx \in t_1^{p} t_2^{q}(\Omega),
	\eeq
	where $(p,q) \in \mathbb{Z}^2$. For the explicit form of $\bfy(\bfx)$, one only needs to substitute the corresponding $\Omega$, $t_1$, $t_2$, $\hat{t}_1$, and $\hat{t}_2$ for the specific pattern in Figures \ref{fig:cones}(c)-(e).

\section{Nondiscrete groups}
\label{sect5}


\subsection{More on helical structures}
As noted in Section \ref{sect1} a structure that is the orbit of a finite set of points under a nondiscrete group of isometries
is not a realistic molecular structure, because nondiscrete groups have accumulation points.
Nevertheless, we argue in the remaining two sections that, properly restricted, these structures are of great interest for
materials science and origami alike.
``Properly restricted'' means that we select the elements of the nondiscrete group in a particular way.  According to the equivalence 
between groups and identical environments presented in Section \ref{sect1}, we cannot select the elements such that, say, each
atom sees the same environment.  However, the examples below show that, by careful selection of the group elements, we
obtain structures in which a) most atoms see the same local environment, or b) each atom sees one of a finite number
of local environments, or c) there are a finite number of local environments and every atom sees one of them, but this number is not fixed, i.e., 
we can have bigger local environments if we allow more of them.  (The latter is a property of a Penrose tiling with atoms at the nodes.) 
Given that all real structures are anyway bounded, these local properties seem 
to us to be quite promising as a basis for the discovery of unusual materials.
\begin{figure}[ht]
\vspace{0mm}
\begin{center}
\subfigure[Partial orbit of a nondiscrete group]{\includegraphics[totalheight=0.1\textheight]{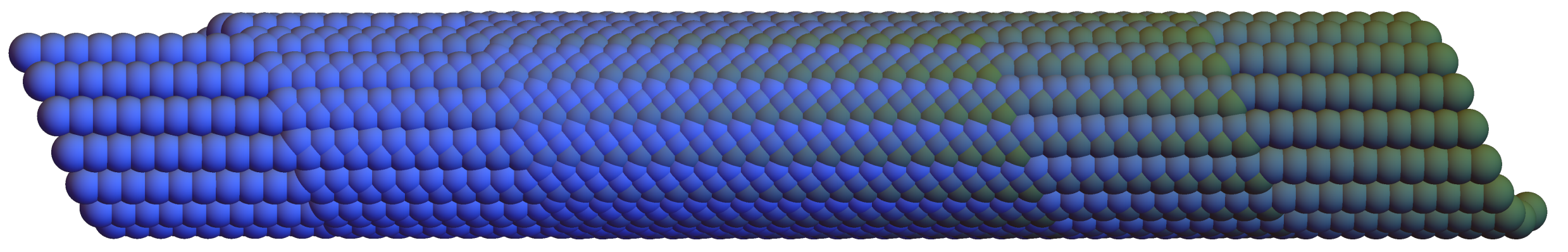}}
\subfigure[Partial orbit of a nondiscrete group with careful selection of powers]{\includegraphics[totalheight=0.1\textheight]{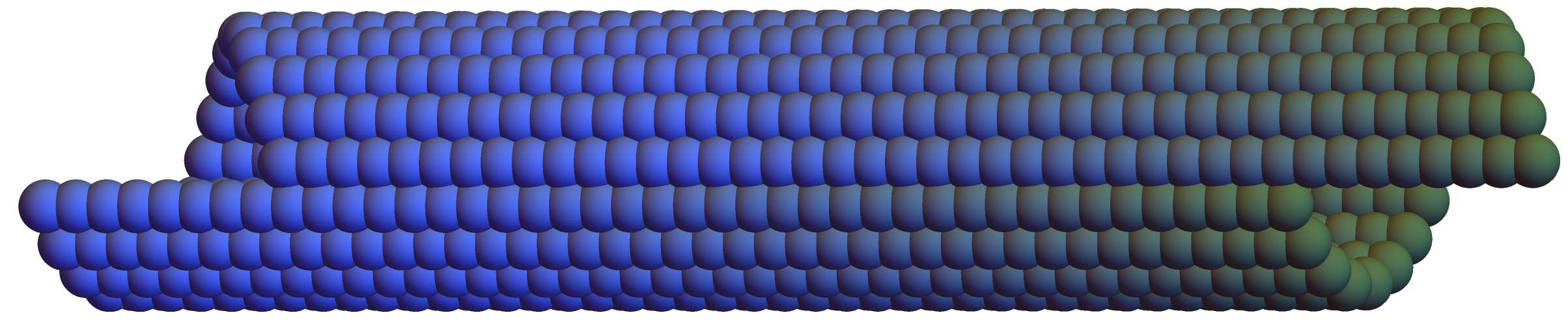}} \\
\vspace{-1mm}
\caption{Orbit $ \{g_1^p g_2^q(\bfx_1): p,q \in \Z \cap \Omega \}$ of a single blue ball at $\bfx_1$ under subsets $\Omega = \Omega_{a,b}$ of a non-discrete 
	helical group defined by regions $\Omega_{a}$ and $\Omega_b$ in $\Z^2$. The shading is based on the value of $q$.  Case (b) shows that $\Omega_b$
	can be chosen so that the atoms not seeing the typical local environment lie on a seam.
	The parameters are (notation of (\ref{defg1g2})): for Case (a), $\psi=\frac{2\sqrt{3}}{9}, \tau m_1=\frac{3}{20},\beta=0, \tau m_2=\frac{1}{4},p=40,q=40$, and for Case (b), $\psi=\frac{\pi}{9}, \tau m_1=\frac{\sqrt{3}}{15},\beta=0, \tau m_2=\frac{1}{4},p=18,q=40$. } 
	\label{fig:nondiscrete_helical}
\end{center}
\vspace{-5mm}
\end{figure}

\begin{figure}[ht]
\vspace{0mm}
\begin{center}
\subfigure[Diatomic structure (with the seam in the back)]{\includegraphics[totalheight=0.13\textheight]{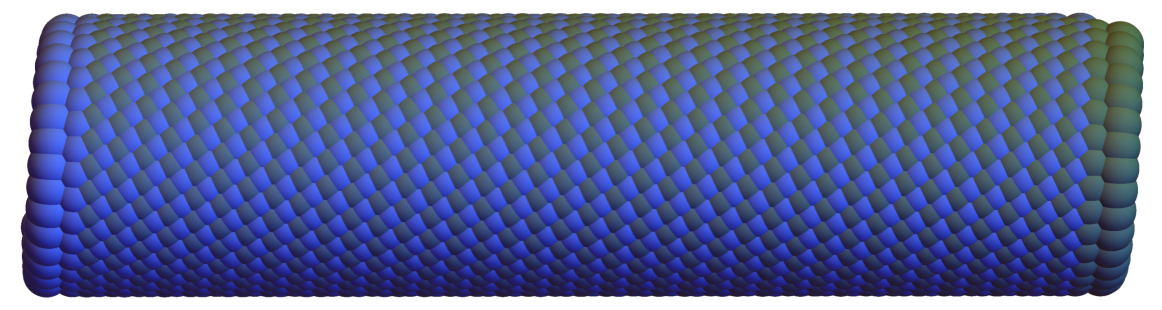}}
\subfigure[Diatomic structure with a helical seam]{\includegraphics[totalheight=0.13\textheight]{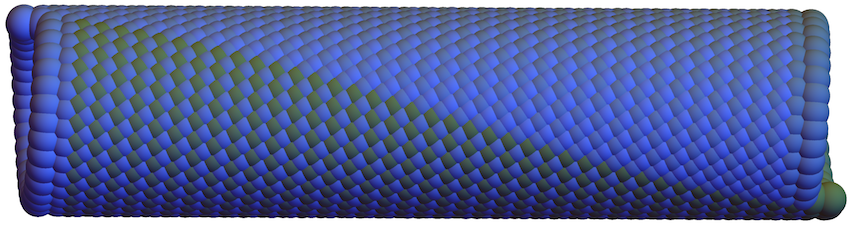}} \\
\vspace{-1mm}
\caption{Diatomic molecular structures obtained as the orbit of a single blue ball under  subsets of nondiscrete helical groups. For Case (a)
the seam is in the back.  Case (b) exhibits a helical seam with the same molecule.  The parameters are $\psi=\frac{\sqrt{3}\pi}{85}, \tau m_1=\frac{\sqrt{3}}{8},\beta=-\frac{123\pi}{2125}, \tau m_2=\frac{1}{100},p=40,q=68$} 
	\label{fig:diatomic}
\end{center}
\vspace{-8mm}
\end{figure}

We begin with the simplest example.  Consider commuting generators $g_1, g_2$ having the form (\ref{defg1g2})  introduced in Section \ref{sect3}
but not satisfying the conditions (\ref{eq:global}) of discreteness.  The structure $ \{g_1^p g_2^q(\bfx_1): p,q \in \Z \}$, with $\bfx_1$ 
not on the axis, generates
points on a cylinder $\calC$ of radius $|\bfz|$ with axis $\bfe$ (Figure 1(a)).     If the discreteness conditions (\ref{eq:global}) fail,  then there are accumulation
points on $\calC$, i.e., $\calG = \{g_1^p g_2^q: p,q \in \Z \}$ is not discrete.   
However, as shown in Figure \ref{fig:nondiscrete_helical}(a), by simply cutting off the powers $p,q$,  large regions of the cylinder become locally objective 
structures  with various size molecules.  In fact, by carefully choosing the powers $p,q$ one can arrange that there is a seam on the 
cylinder parallel to the axis $\bfe$, and each atom away from this seam sees the
same local environment, Figure \ref{fig:nondiscrete_helical}(b).   And, curiously, the atoms right next to the seam (on one side) also all see the 
same environment.  One can arrange also that the seam is helical, Figure \ref{fig:diatomic}(b).

A close examination of Figure \ref{fig:nondiscrete_helical}(a) reveals locally objective molecular structures with molecules of different size.  By selecting the
powers $p,q$ suitably, one can also make a uniform 
molecule.  Figure (\ref{fig:diatomic}) shows a case with a diatomic molecule.  Again, necessarily, 
there is a seam, which is chosen to be helical in Figure \ref{fig:diatomic}(b). 


The nondiscrete groups here offer a lot of additional freedom on the structure of the molecule and its placement
with regard to its neighbors, at the expense of
a seam.  For example, a much enlarged set of lattice parameters becomes possible that would not be possible with a helical objective atomic structure. 
Since we have no idea what are the nondiscrete groups of isometries, we do not know the scope of these methods at this time.  So, we
confine attention to examples.
A familiar biological example of a structure of the type shown in Figure \ref{fig:nondiscrete_helical}(b) is the microtubule.  In fact, 
it is argued in  \cite{microtubule_seam} that the axial seam of the microtubule is functional and aids in assembly and disassembly of the microtubule.


\subsection{Viruses and quasicrystals}
\label{sect6}

In this section we explain a relation between the use of nondiscrete groups and known methods
of describing the structures of animal viruses and quasicrystals.

\subsubsection{Virus structure}

Reidun Twarock, Thomas Keef and collaborators \cite{keef2009, twarock2004, twarock2010, keef2013, twarock2014} developed a way of looking at the structure of 
icosahedral viruses, especially of the families Papovaviridae and Nodaviridae,  that generalizes the celebrated ideas of 
Caspar and Klug \cite{caspar_62}.   Of interest here is their method of modeling
the positions of the spikes on the virus, and the arrangement of molecules below the spikes, as structures obtained by affine extensions
of the icosahedral group\footnote{Incidentally, the results do not apply to the Covid-19 virus.  In that case the proteins, including the spike (S) protein, are glycoproteins in a viral envelope, and therefore do not occupy such well defined positions (personal communication, Reidun Twarock).}.  The locations of the spikes, and their terminal molecules, are critical for the ability of the virus to avoid recognition by the host, and thus the work has medical 
implications.  Here we show that the method of affine extensions also corresponds to the judicious selection of powers of a nondiscrete isometry 
group with its locally identical environments.



\begin{figure}[hb!]
	\center
	\includegraphics[width=0.8\textwidth]{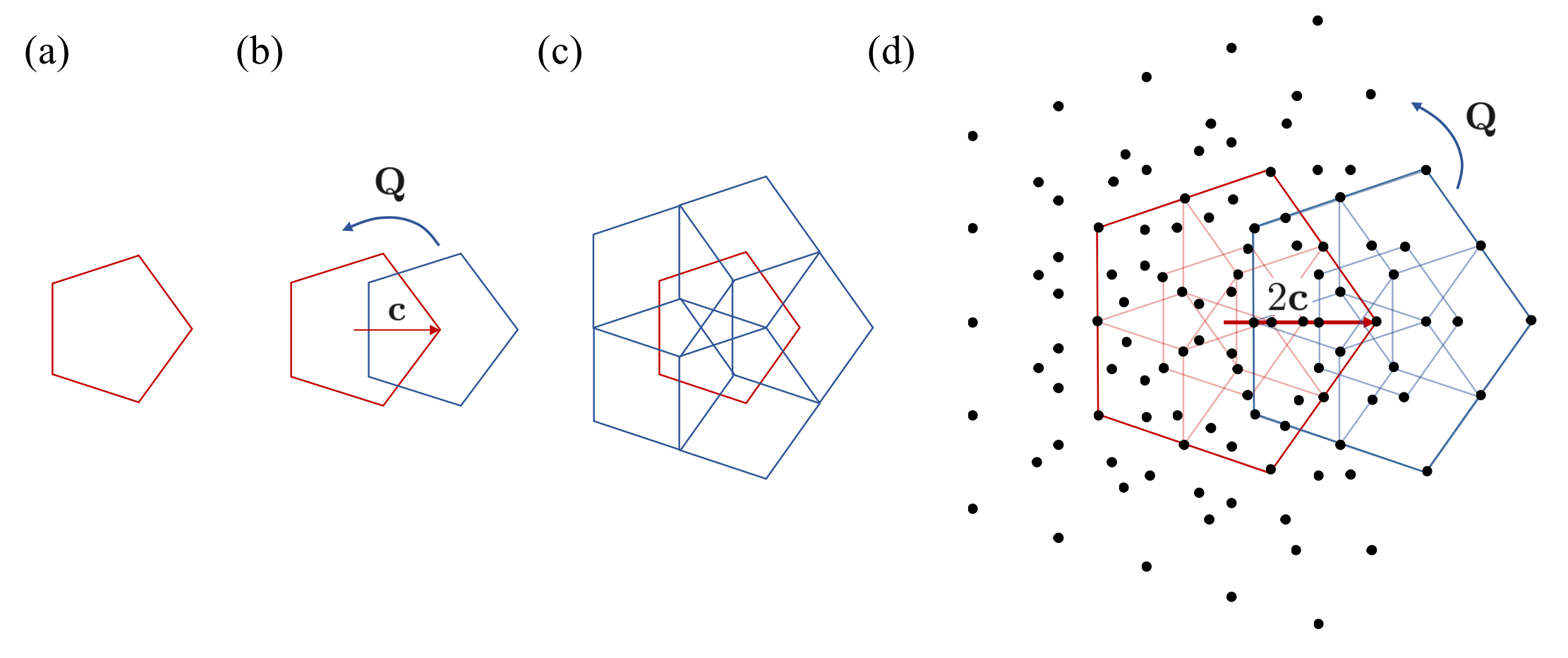}
	\vspace{-5mm}
	\caption{A lattice-like point set constructed by a cyclic group and a translation operator.} 
	\label{fig:2D_example}
\end{figure}

To appreciate this assertion, we begin with a simplified 2D model of Keef and Twarock \cite{twarock2008, keef2009} that explains their idea, Figure \ref{fig:2D_example}.  In this example the analog of the icosahedral group is the cyclic group of 5-fold rotations. As illustrated in Figure \ref{fig:2D_example} the point set of interest is obtained by taking the orbit of the pentagon under the group generated by the two isometries
\beq
\hat g_1=(\bfI|\bfc),\; \hat g_2=(\bfQ|\mathbf{0}). 
\eeq
where $\bfQ$ is a $2\pi/5$ rotation with axis perpendicular to the plane. Recall that composition of mappings corresponds to the group product of isometries and notice that the sequence of operations pictured in Figure  \ref{fig:2D_example} is  $\dots \hat{g}_2^j \hat{g}_1 \hat{g}_2^i \hat{g}_1 (\pentagon)\ i,j = 1,\dots, 5$.

As remarked by Keef and Twarock, if continued indefinitely, this set is not discrete; accumulation points lie on certain radial lines.  To see the nondiscreteness easily, we observe that  (i) $g_i := \hat g_2^{i}\hat g_1\hat g_2^{-i}=(\bfI|\bfQ^i\bfc), i \in \Z$, is a translation, and that (ii) for any $k \in \Z$,
$ \hat{g}_2^{k} (\pentagon) =  \pentagon$.  Therefore, the point set of Figure \ref{fig:2D_example} can be written
\beq
\dots \hat{g}_2^j \hat{g}_1 \hat{g}_2^i \hat{g}_1 (\pentagon)  =  \dots (\hat{g}_2^j \hat{g}_1 \hat{g}_2^{-j}) (\hat{g_2}^{j+i} \hat{g}_1 \hat{g}_2^{-(j+i)})(\pentagon)  = \dots g_{ j}  g_{i+j} (\pentagon)     \quad   i,j = 1,\dots, 5   \label{pentagroup}
\eeq

This formula, which can be continued indefinitely to the left, shows that the point set of Figure \ref{fig:2D_example}  can be generated
by selecting elements from the Abelian group of translations generated by the five elements
\beq
g_1=(\bfI|\bfQ\bfc),\;g_2=(\bfI|\bfQ^2\bfc),\; g_3=(\bfI|\bfQ^3\bfc),\; g_4=(\bfI|\bfQ^4\bfc),\;g_5=(\bfI|\bfc).
\eeq
Observing  that $g_2g_3=(\bfI|\bfQ^2\bfc+\bfQ^3\bfc)=(\bfI|-2\cos(\frac{\pi}{5})\bfc)=(\bfI|-(\frac{\sqrt{5}+1}{2})\bfc)$, is an irrational translation,
we see that this group is nondiscrete.  The selection of powers is given in (\ref{pentagroup}).

Like the other examples in this section, by restricting the number of iterations in (\ref{pentagroup}), they arrive at a point system that, in the corresponding 3D case, is in remarkable agreement with the structure of the spikes and the underlying molecules.

\begin{figure}[ht]
	\center
	\includegraphics[width=\textwidth]{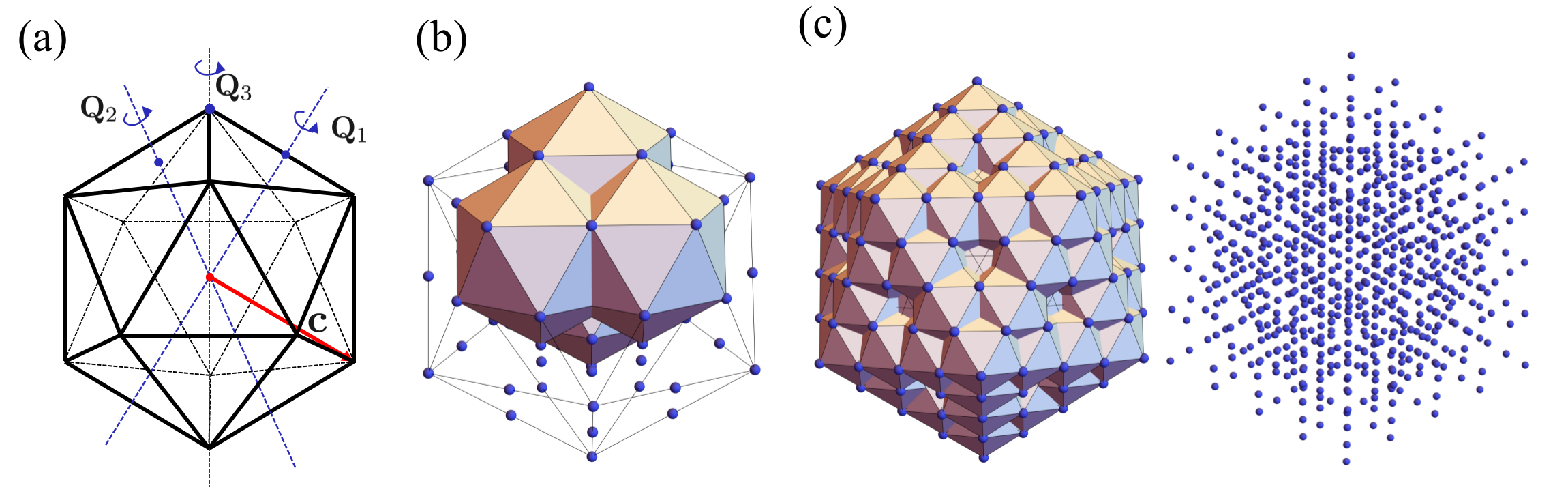}
	\caption{(a) Starting configuration showing the icosahedral group and an initial translation. Point sets in the first (b) and the second (c) iterations. } 
	\label{fig:icosahedron}
\end{figure}

In the case of the real virus, the 2D starting configuration is replaced by a 3D structure, such as the icosahedron,  dodecahedron, or icosidodecahedron \cite{keef2009, keef_13}. The point group $\bfC_5$ and translation along the pentagon vertex (Figure \ref{fig:2D_example}(b)) are changed to the icosahedral group $\calI$ and translation along the 5-, 3- or 2-fold axis of icosahedral symmetry, respectively. Taking the starting configuration as an icosahedron and translation along the 5-fold axis as an example, the 3D point set is obtained by taking the orbit of the icosahedron under the group generated by the four isometries
\beq
\hat g_1=(\bfI|\bfc),\;\hat g_2=(\bfQ_1|\mathbf{0}),\;\hat g_3=(\bfQ_2|\mathbf{0}),\;\hat g_4=(\bfQ_3|\mathbf{0}).
\eeq
$\bfQ_1,\bfQ_2$ and $\bfQ_3$ are rotations with 2-, 3- and 5-fold rotational axes, respectively, as illustrated in Figure $\ref{fig:icosahedron}$(a). The sequence of operations to extend the point set is $\dots\hat g_4^{j_3}\hat g_3^{j_2}\hat g_2^{j_1}\hat g_1\hat g_4^{i_3}\hat g_3^{i_2}\hat g_2^{i_1}\hat g_1(\begin{tikzpicture} \newdimen\r;\r=1.35mm;\draw  [thick,black!80](-30:\r) \foreach \x in {30,90,150,210,270,330} {-- (\x:\r)};\end{tikzpicture}),$ $i_1,j_1=1,2;\ i_2,j_2=1,2,3;\ i_3,j_3=1,\dots,5$, and
\begin{tikzpicture} \newdimen\r;\r=1.35mm;\draw  [thick,black!80](-30:\r) \foreach \x in {30,90,150,210,270,330} {-- (\x:\r)};\end{tikzpicture} 
 represents the icosahedron.


Similarly, we can find that (i) $g_i:=\hat g_4^{i_3}\hat g_3^{i_2}\hat g_2^{i_1}\hat g_1\hat g_2^{-i_1}\hat g_3^{-i_2}\hat g_4^{-i_3}=(\bfI|\bfQ_1^{i_1}\bfQ_2^{i_2}\bfQ_3^{i_3}\bfc)$ gives a translation, and that (ii) $\hat g_{2,3,4}^{k}$
(\begin{tikzpicture}\newdimen\r; \r=1.35mm; \draw [thick,black!80](-30:\r) \foreach \x in {30,90,150,210,270,330} {-- (\x:\r)};\end{tikzpicture})
=
\begin{tikzpicture} \newdimen\r;\r=1.35mm;\draw  [thick,black!80](-30:\r) \foreach \x in {30,90,150,210,270,330} {-- (\x:\r)};\end{tikzpicture}. Therefore
the same point set is given by
\beqs
&&\dots\hat g_4^{j_3}\hat g_3^{j_2}\hat g_2^{j_1}\hat g_1\hat g_4^{i_3}\hat g_3^{i_2}\hat g_2^{i_1}\hat g_1(\begin{tikzpicture} \newdimen\r;\r=1.35mm;\draw  [thick,black!80](-30:\r) \foreach \x in {30,90,150,210,270,330} {-- (\x:\r)};\end{tikzpicture})\nonumber\\
&=&\dots(\hat g_4^{j_3}\hat g_3^{j_2}\hat g_2^{j_1}\hat g_1\hat g_2^{-j_1}\hat g_3^{-j_2}\hat g_4^{-j_3})(\hat g_4^{i_3+j_3}\hat g_3^{i_2+j_2}\hat g_2^{i_1+j_1}\hat g_1\hat g_2^{-(i_1+j_1)}\hat g_3^{-(i_2+j_2)}\hat g_4^{-(i_3+j_3)})(\begin{tikzpicture} \newdimen\r;\r=1.35mm;\draw  [thick,black!80](-30:\r) \foreach \x in {30,90,150,210,270,330} {-- (\x:\r)};\end{tikzpicture} )\nonumber\\
&=&\dots g_jg_{i+j}(\begin{tikzpicture} \newdimen\r;\r=1.35mm;\draw  [thick,black!80](-30:\r) \foreach \x in {30,90,150,210,270,330} {-- (\x:\r)};\end{tikzpicture} ) \quad i,j=1,\dots,30   \label{icosahedrongroup}
\eeqs

Thus, one can obtain the point set by selecting elements from the Abelian group of translations generated by the twelve elements (by eliminating repeated elements in $g_i,i=1,\dots,30$)
\beqs
g_1=(\bfI|\bfc),\;g_2=(\bfI|\bfQ_1\bfc),\;g_3=(\bfI|\bfQ_2\bfc),\;g_4=(\bfI|\bfQ_2^2\bfc),\;
g_5=(\bfI|\bfQ_2\bfQ_1\bfc),\;g_6=(\bfI|\bfQ_2^2\bfQ_1\bfc),\quad\quad\\ \label{icosahedron_generator}
g_7=(\bfI|\bfQ_3\bfc),\;g_8=(\bfI|\bfQ_3^2\bfc),\;g_9=(\bfI|\bfQ_3^3\bfc),\;g_{10}=(\bfI|\bfQ_3^2\bfQ_1\bfc),\;g_{11}=(\bfI|\bfQ_3^3\bfQ_1\bfc),\;g_{12}=(\bfI|\bfQ_3^4\bfQ_1\bfc)\nonumber
\eeqs

Notice that $g_6g_{10}=(\bfI|\bfQ_2^2\bfQ_1\bfc+\bfQ_3^2\bfQ_1\bfc)=(\bfI|\frac{\sqrt{50+10\sqrt{5}}}{5}|\bfc|\bfe)$ is an irrational translation ($\bfe$ is the rational axis of $\bfQ_1$), so the Abelian group generated by ($\ref{icosahedron_generator}$) is nondiscrete. See the selection of powers in ($\ref{icosahedrongroup}$).

\subsubsection{Quasicrystals}
Therefore, by cutting off powers or selecting powers of generators in non-discrete groups, one can obtain realistic structures, i.e., no accumulated points or patterns in the structures.  This provides a good way to build discrete structures with non-discrete groups. Here is an example of using non-discrete groups to construct origami structures. The non-discrete group we use is the one in 2D virus case with the following five generators
\beq
g_1=(\bfI|\bfQ\bfc),\;g_2=(\bfI|\bfQ^2\bfc),\;g_3=(\bfI|\bfQ^3\bfc),\;g_4=(\bfI|\bfQ^4\bfc),\;g_5=(\bfI|\bfc).
\eeq
where $\bfQ$ is a rotation with an angle of $2\pi/5$ and $\bfc$ is a translation. Different from the virus case
 \begin{figure}[ht]
	\center
	\includegraphics[width=1\textwidth]{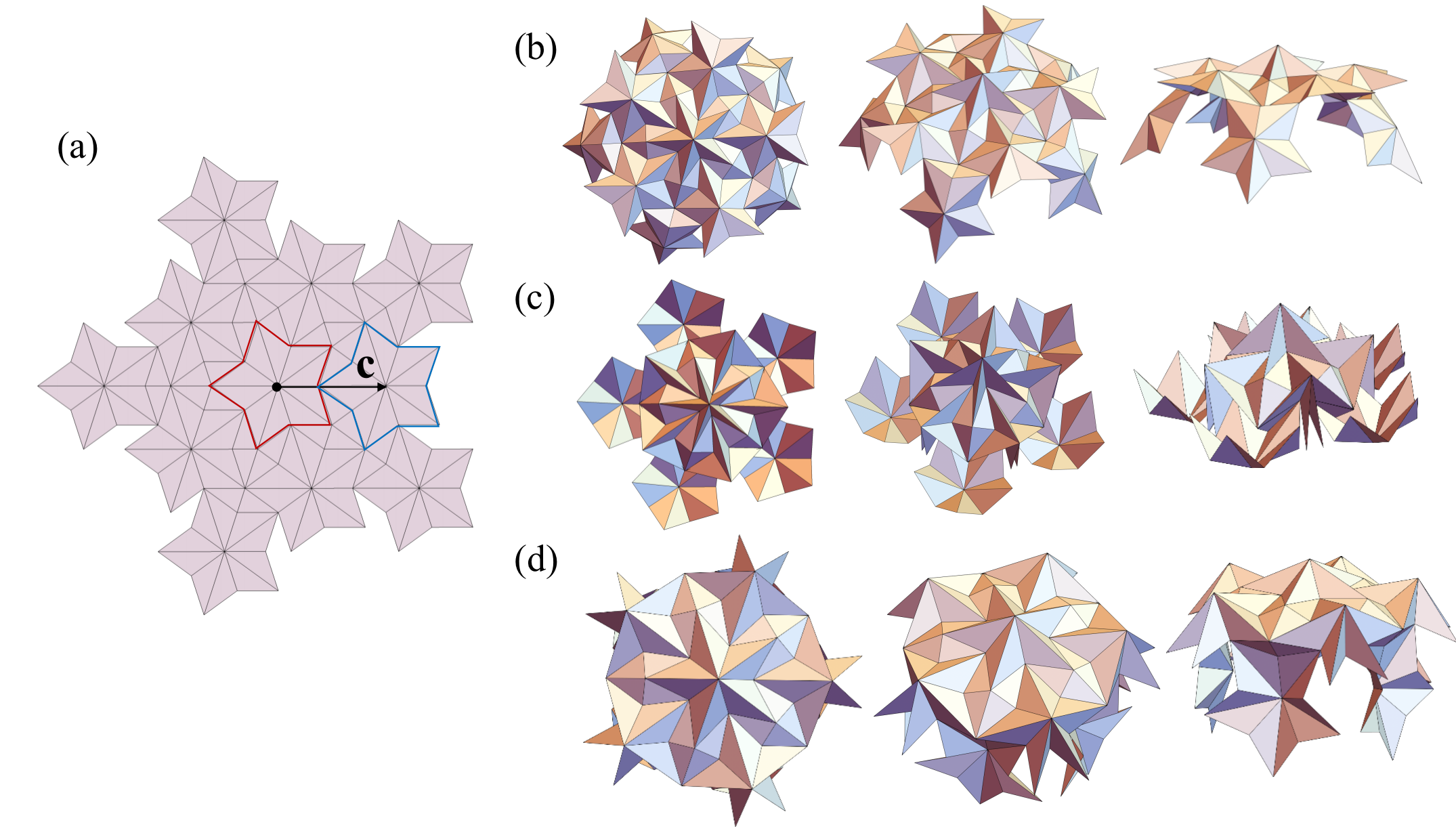}
	\caption{(a) Crease pattern of origami generated by the subset of non-discrete group, in which a star-like reference
configuration and an initial translation are shown. (b-d) Three examples of folding ways with different mountain-valley assignments. Each example is displayed from three viewpoints.} 
	\label{fig:origami}
\end{figure}
 in which each pentagon intersects with its neighbors, we use a star-like reference configuration (one can change into other shapes with five-fold symmetry) and the length of the selected translation is the length of the star along the axis of symmetry, which allows the units to be discrete as shown in Fig.$\ref{fig:origami}$(a). The star is composed of five rhombuses with the internal angle of $2\pi/5$. Here we choose the powers of generators such that some stars see the same environment in the finite environments. Setting the reference star $\bfX$ as the identity and using the cyclic permutation $\hat\sigma=\left( \begin{matrix}
1 & 2 & 3 & 4 & 5 \\
2 & 3 & 4 & 5 &1 
\end{matrix} \right)$, i.e., $\hat\sigma(1)=2, \hat\sigma(2)=3, \hat\sigma(3)=4, \hat\sigma(4)=5$, and $\hat\sigma(5)=1,$ we give the orbit of the star as follows:
\beq
\{\bfX+p\bfQ^{\hat\sigma(i)}\bfc+q\bfQ^{\hat\sigma^2(i)}\bfc:p,q\in \{ 0, 1,2 , \ldots\} , i=1,...,5\}
\eeq
where $\hat\sigma^2(i)=\hat\sigma(\hat\sigma(i))$. The corresponding subset of the non-discrete group is 
\beq
\widetilde{\calG}=\{g_{\hat\sigma(i)}^pg_{\hat\sigma^2(i)}^q: p,q\in \{ 0, 1,2 , \ldots\} , i=1,...,5\}
\eeq
When $p=q=0$, $g_{\hat\sigma(i)}^pg_{\hat\sigma^2(i)}^q$ gives the identity. Under this subset, the plane can be tessellated aperiodically with two shapes: the thick rhombus in the stars and the thin rhombuses with the internal angle of $\pi/5$. Since the generators in the subset show a cyclic permutation, the whole pattern shows five-fold symmetry.

In Figure $\ref{fig:origami}$(a), the crease pattern is obtained by restricting the powers $p,q=0,1$, which corresponds to $\{g_{\hat\sigma(i)}^pg_{\hat\sigma^2(i)}^q($\ding{73}$):p,q=0,1, i = 1,\ldots, 5\}$. The crease pattern has 60 degrees of freedom (DOF) for rigid folding. We add the symmetry in the folding process that reduces the DOF to 6. Figure $\ref{fig:origami}$(b) shows two examples of folding ways with different mountain-valley assignments. There are three different environments in the crease pattern, which are the environments seen by (i) the identity $\bfI$, (ii) the stars $\{g_{\hat\sigma(i)}^pg_{\hat\sigma^2(i)}^q($\ding{73}$):p=1,q=0, i = 1,\ldots,5\}$, and (iii) the stars $\{g_{\hat\sigma(i)}^pg_{\hat\sigma^2(i)}^q($\ding{73}$):p=1,q=1, i = 1,\ldots, 5\}$, respectively. During the folding process, one can see that the stars that see the same environments show the same folding configurations.

\appendix

\section{Appendix}

\subsection{Proof of Proposition \ref{prop1}}

\noindent Proof. If $G$ is not discrete,
then, for each point $\bfx \in \calS$, we claim that $G$
contains an infinite number of isometries that fix $\bfx$.
To this end suppose that $G$ is the isometry group of $\calS$
and suppose that $\calS$ is discrete but $G$ is not discrete.
Then there is a point
$\bfy \in \R^3$ and an infinite sequence of elements
$g_i = (\bfQ_i|\bfc_i) \in G$
such that $g_i(\bfy) \to \bfz$ with $g_i(\bfy)$ distinct.
Let $\bfx \in \calS$.   Since
$\calS$ is discrete, $g_i(\bfx)$ cannot contain a convergent
sequence of distinct points, that is, either a) $|g_i(\bfx)| \to
\infty$ or b) the range of $g_i(\bfx)$ consists of a finite number
of points. We claim that (a) cannot occur. On the contrary, if (a)
holds, then we have for some $\bfy \ne \bfx \in \calS$
$|g_i(\bfx)| \to \infty$ and $g_i(\bfy) \to \bfz$.
Thus $|g_i(\bfx) - g_i(\bfy)| \to \infty$.  But since the $g_i$
are isometries
$|g_i(\bfx) - g_i(\bfy)|  = |\bfx - \bfy|$, a contradiction.
Thus, we have the remaining alternative (b),  the range of
$g_i(\bfx)$ consists of a finite number
of points.  One of these points must be taken on infinitely
often, so we must have, for a subsequence (not relabeled),
$g_i(\bfx) = \bfb,\, i = 1, 2, \dots$.  Thus, $\bfb \in \calS$.
If $\bfb$ is not already equal to $\bfx$, we can find a $g \in G$ such
that $g (\bfb) = \bfx$.  Then we note that $g g_i (\bfx) = \bfx,\,
 i = 1, 2, \dots$. Since they fix a point, the infinite
 number of distinct elements $g g_i$ represent pure orthogonal 
 transformations about that point.

For each $\bfx \in \calS$ let
 $G_{\bfx} = \{ g \in G: g(\bfx) = \bfx \}$.  The above shows that
 each $G_{\bfx}$ is an infinite group.  We claim
 that if $\calS$ contains more than one point, then all points of $\calS$
 lie on a line, and this line is invariant
 under $G_{\bfx}$.
 Suppose $\bfx' \in \calS,
 \bfx' \ne \bfx$.  Since $G_{\bfx}$ is an infinite group of isometries
 fixing $\bfx$, then $G_{\bfx}(\bfx')$ is a collection
of points of $\calS$  on a sphere centered at $\bfx$ with radius
$|\bfx' - \bfx|$.  Since $\calS$ is discrete, it follows that
 an infinite number of
elements of $G_{\bfx}$ must map $\bfx'$ to some $\bfx'' \in \calS$.
Let $J_{\bfx} = \{ g \in G_{\bfx}: g(\bfx') = \bfx'' \}$ be this
infinite set of elements.  Let $a \in J_{\bfx}$ and
let $J_{\bfx}' = \{g a^{-1}:
g \in J_{\bfx} \}$.  If $g \in J_{\bfx}'$ then clearly $g(\bfx'') = \bfx'' $.
This shows that there are an infinite number of distinct isometries
in $G_{\bfx}$ that also fix $\bfx'' \ne \bfx$.  We claim that there
are in fact an infinite sequence of proper rotations in $G_{\bfx}$ with
this property.  If that were not true, then there would necessarily
 be an infinite number of improper rotations  $g_i \in G_{\bfx}$,
 $g_i = (\bfQ_i|(\bfI - \bfQ_i)\bfx)$, that satisfy $\bfQ_i\bfe
 = \bfe$, $\bfe = \bfx''-\bfx$.  But then, $g_ig_1^{-1}$ are
 an infinite number of distinct proper rotations that fix
 both $\bfx$ and $\bfx''$.
 
Clearly, then, all elements of $\calS$ must be on the line
$\bfx + \lambda \bfe, \, \lambda \in \R$. For, if $\bfz$
is not on this line, then the infinite number of distinct proper
rotations in $G_{\bfx} \subset G$ would map $\bfz$ to an infinite
number of distinct points on a circle.

It follows that $\calS$ is then a
one-dimensional objective atomic structure. Obviously, this
includes structures with one or two points.  If $\calS$ has
at least three points $\bfx_1, \bfx_2, \bfx_3$, consecutively
along a line, then using the concept of an objective structure
(identical environments) it is clear that $\bfx_3$ must have
a neighbor $\bfx_4 = \bfx_3 + (\bfx_2 - \bfx_1)$.  Continuing in
this fashion, we generate one of the two possibilities given
in the statement of the proposition.

\vspace{2mm}
\noindent {\bf Acknowledgment}  The authors thank Reidun Twarock for helpful comments.  
The authors thank the Isaac Newton Institute for Mathematical Sciences for support during the 
program ``The mathematical design of new materials'' supported by EPSRC grant  EP/R014604/1. The
residence of RDJ there was supported by a Simons Fellowship. 
RDJ and HL also benefited from the support of ONR (N00014-18-1-2766), MURI (FA9550-18-1-0095), and
a Vannevar Bush Faculty Fellowship, and all authors acknowledge the support of the MURI project FA9550-16-1-0566.

\bibliographystyle{plain}
\bibliography{O_ms.bib}

\end{document}